\documentclass[aps,prc,twocolumn,showpacs]{revtex4}

\usepackage{graphicx}
\usepackage{longtable}
\usepackage{amssymb,amsmath}
\usepackage{subfigure}

\begin{document}
\pagenumbering{arabic}


\title{Nuclear spins, magnetic moments and quadrupole moments of Cu isotopes from $N=28$ to $N=46$: probes for core polarization effects}


\author{P. Vingerhoets$^1$, K.T. Flanagan$^{1,2}$, M. Avgoulea$^1$, J. Billowes$^3$, M.L. Bissell$^1$, K. Blaum$^4$, B.A. Brown$^5$, B. Cheal$^3$, M. De Rydt$^1$, D.H. Forest$^6$, Ch. Geppert$^{7,10}$, M. Honma$^8$, M. Kowalska$^{9}$, J. Kr\"{a}mer$^{10}$, A. Krieger$^{10}$, E. Man\'{e}$^3$, R. Neugart$^{10}$, G. Neyens$^1$, W. N\"{o}rtersh\"{a}user$^{7,10}$, T. Otsuka$^{11}$, M. Schug$^4$, H.H. Stroke$^{12}$, G. Tungate$^6$, D.T. Yordanov$^{1,4}$ }

\affiliation{\vspace{0.4cm}$^1$Instituut voor Kern- en Stralingsfysica, K.U. Leuven, B-3001 Leuven, Belgium}
\affiliation{$^2$IPN Orsay, F-91940 Orsay Cedex, France}
\affiliation{$^3$School of Physics and Astronomy, The University of Manchester, Manchester, M13 9PL, United Kingdom}
\affiliation{$^4$Max-Planck-Institut f\"{u}r Kernphysik, D-69117 Heidelberg, Germany}
\affiliation{$^5$National Superconducting Cyclotron Laboratory and Department of Physics and Astronomy, Michigan State University, East Lansing, Michigan 48824-1321, USA}
\affiliation{$^6$School of Physics and Astronomy, The University of Birmingham, Birmingham, B15 2TT United Kingdom}
\affiliation{$^7$GSI Helmholtzzentrum f\"{u}r Schwerionenforschung GmbH, D-64291 Darmstadt, Germany}
\affiliation{$^8$Center for Mathematical Sciences, University of Aizu, Tsuruga, Ikki-machi, Aizu-Wakamatsu, Fukushima 965-8580, Japan}
\affiliation{$^{9}$Physics Department, CERN, CH-1211 Geneva 23, Switzerland}
\affiliation{$^{10}$Institut f\"{u}r Kernchemie, Johannes Gutenberg-Universit\"{a}t Mainz, D-55128 Mainz, Germany}
\affiliation{$^{11}$RIKEN, Hirosawa, Wako-shi, Saitama 351-0198, Japan}
\affiliation{$^{12}$Department of Physics, New York University, New York, New York 10003, USA}



\date{\today}

\begin{abstract}
Measurements of the ground-state nuclear spins, magnetic and quadrupole moments of the copper isotopes from $^{61}$Cu up to $^{75}$Cu are reported. The experiments were performed at the ISOLDE facility, using the technique of collinear laser spectroscopy. The trend in the magnetic moments between the $N=28$ and $N=50$ shell closures is reasonably reproduced by large-scale shell-model calculations starting from a $^{56}$Ni core.  The quadrupole moments reveal a strong polarization of the underlying Ni core when the neutron shell is opened, which is however strongly reduced at $N=40$ due to the parity change between the $pf$ and $g$ orbits. No enhanced core polarization is seen beyond $N=40$. Deviations between measured and calculated moments are attributed to the softness of the $^{56}$Ni core and weakening of the $Z=28$ and $N=28$ shell gaps.
\end{abstract}


\maketitle

\section{\label{intro}Introduction}

A key question in nuclear structure research is the persistence of so-called magic numbers when moving away from stability. Nuclei near closed shells are important testing grounds for shell-model theories and have therefore attracted considerable experimental and theoretical research interest. A particularly interesting region is around the magic number of $Z=28$ protons, as it ranges from the doubly-magic $^{56}$Ni on the neutron-deficient side of the nuclear chart towards the neutron-rich doubly magic $^{78}$Ni, 14 isotopes away from stability.  Furthermore, it includes the semi-magic sub-shell closure at $N=40$, which is related to the parity change between the $pf$ shell and the $g_{9/2}$ orbital \cite{Mue99, Sor02, Lan03, Gue07}. The nickel region has been investigated extensively in the last decade both theoretically and experimentally. On the neutron-deficient side, where protons and neutrons occupy the negative-parity $pf$ orbits, it has been shown experimentally~\cite{Ken01,Coc09} and theoretically~\cite{Hon02,Hon04} that the $^{56}$Ni core is rather soft.  Excitations of protons and neutrons across the $N=Z=28$ shell closure from the $f_{7/2}$ orbital into the higher $pf$ orbits are needed to reproduce the magnetic moments of the $2^+$ states in the even $^{58-64}$Ni isotopes~\cite{Ken01}. Of special interest in the region are the ground-state properties of the copper isotopes, which are dominated by a single proton coupling to the underlying nickel core. Magnetic moment measurements of odd-A Cu isotopes have been performed over a very broad range, from $N=28$ up to $N$=46~\cite{Rag89,Rik00, Rik00b,Sto08,Geo02,Coc09,Fla09}. On the neutron-deficient side, excitations of nucleons from the $f_{7/2}$ orbit across the spin-orbit magic numbers $N=Z=28$ are needed to reproduce the observed moments for the $\pi p_{3/2}$ ground states~\cite{Coc09,Sto08}, as shown by the calculations in the full $pf$-shell \cite{Hon04}.  On the neutron-rich side, the inversion of the ground-state structure from $\pi p_{3/2}$-dominated to $\pi f_{5/2}$-dominated was established by the measured ground-state spins of $^{73,75}$Cu~\cite{Fla09}.  In this region the strong interaction between the $f_{5/2}$ protons and the $g_{9/2}$ neutrons plays a crucial role \cite{Ots05,Ots10}.  This has been taken into account in two effective shell-model interactions which span the $p_{3/2}\,f_{5/2}\,p_{1/2}\,g_{9/2}$ model space (often abbreviated as the $f5pg9$ model space) based on a $^{56}$Ni core \cite{Hon09}, and which reproduce the odd $^{69-75}$Cu magnetic moments fairly well~\cite{Fla09}. However, an increasing deviation was observed towards $^{73}$Cu, which is suggested to be due to missing proton excitations across $Z=28$.  Indeed, calculations starting from a $^{48}$Ca core can reproduce nearly perfectly the experimental odd-A Cu magnetic moments beyond $N=40$ \cite{Sie10}.

In this paper, we report on the measured magnetic moments and quadrupole moments of isotopes between $^{61}$Cu and $^{75}$Cu.  The moments and spins of nuclear ground and long-lived isomeric states were determined using high-resolution collinear laser spectroscopy. With this method, each of these observables could be deduced in a model-independent way from the measured hyperfine spectra in the \nolinebreak{$3d^{10}4s\,\,\,^2S_{1/2} \rightarrow 3d^{10}4p\,\,\, ^2P_{3/2}$} transition for atomic copper. A review of recent developments on laser spectroscopy can be found in \cite{Che10b}. In order to extend the measurements towards the neutron-deficient and neutron-rich sides of the valley of stability, the sensitivity of the optical detection has been enhanced by more than two orders of magnitude, by using a bunched ion beam produced with the recently installed gas-filled linear Paul trap (ISCOOL)~\cite{Fra08,Man09}. The data are compared to shell-model calculations using a $^{56}$Ni core.

\section{\label{exp}Experimental set-up and procedure}

\begin{figure}[htb]
\includegraphics[scale = 0.35]{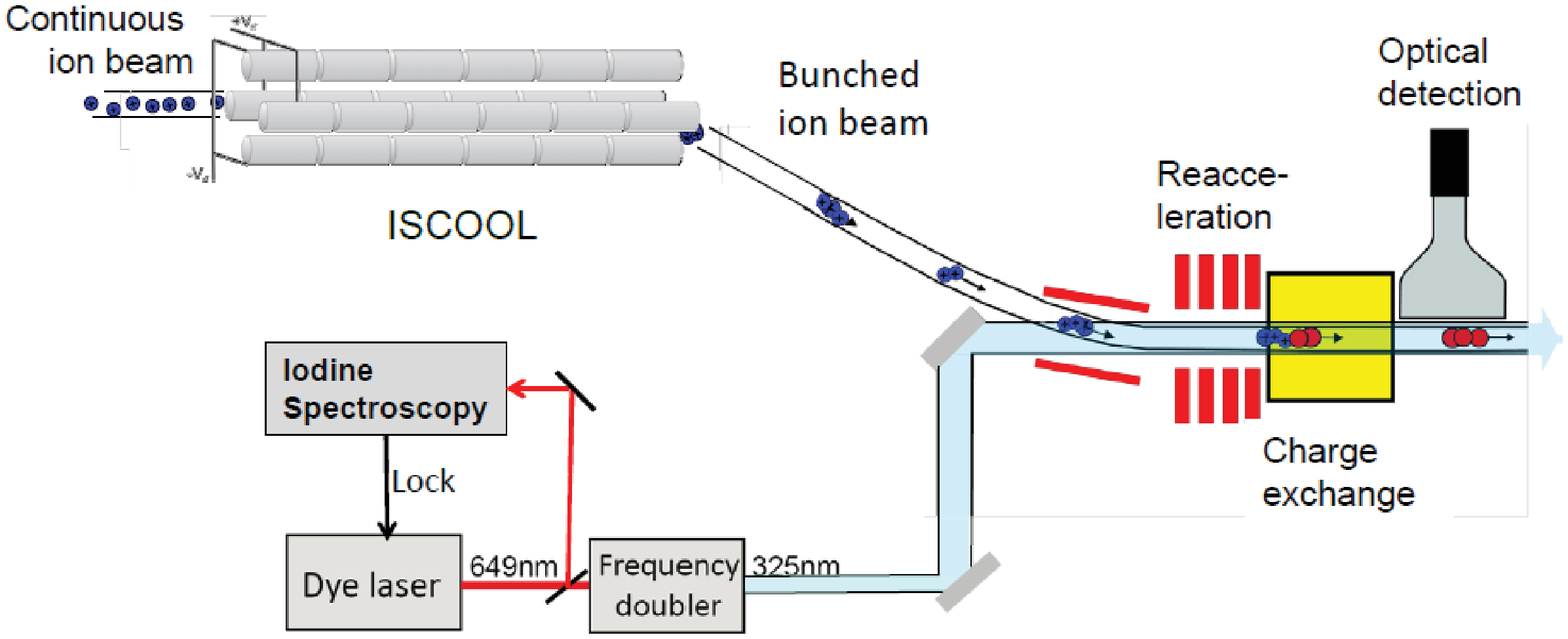}
\caption{\label{setup}(Color online) Experimental setup of the COLLAPS collinear laser spectroscopy beamline.}
\end{figure}

\begin{table}\caption{\label{experiments}Overview of the isotopes measured during the three COLLAPS beamtimes. The on-line commissioning of the ISCOOL cooler/buncher behind the HRS allowed investigation of more exotic isotopes in 2008.}
\begin{ruledtabular}
\begin{tabular}{c|c|c}
year&separator&Isotopes addressed\\
\hline
2006&GPS&$^{63,64,65,66,67,68^g,68^m,69,70^g}$Cu\\
2007&GPS&$^{62,63,65,67,69,70^g,71,72}$Cu\\
2008&HRS&$^{61,65,68^g,68^m,70^g,70^{m1},70^{m2},71,72,73,74,75}$Cu\\
\end{tabular}
\end{ruledtabular}
\end{table}

\begin{figure}
\includegraphics[scale = 0.35]{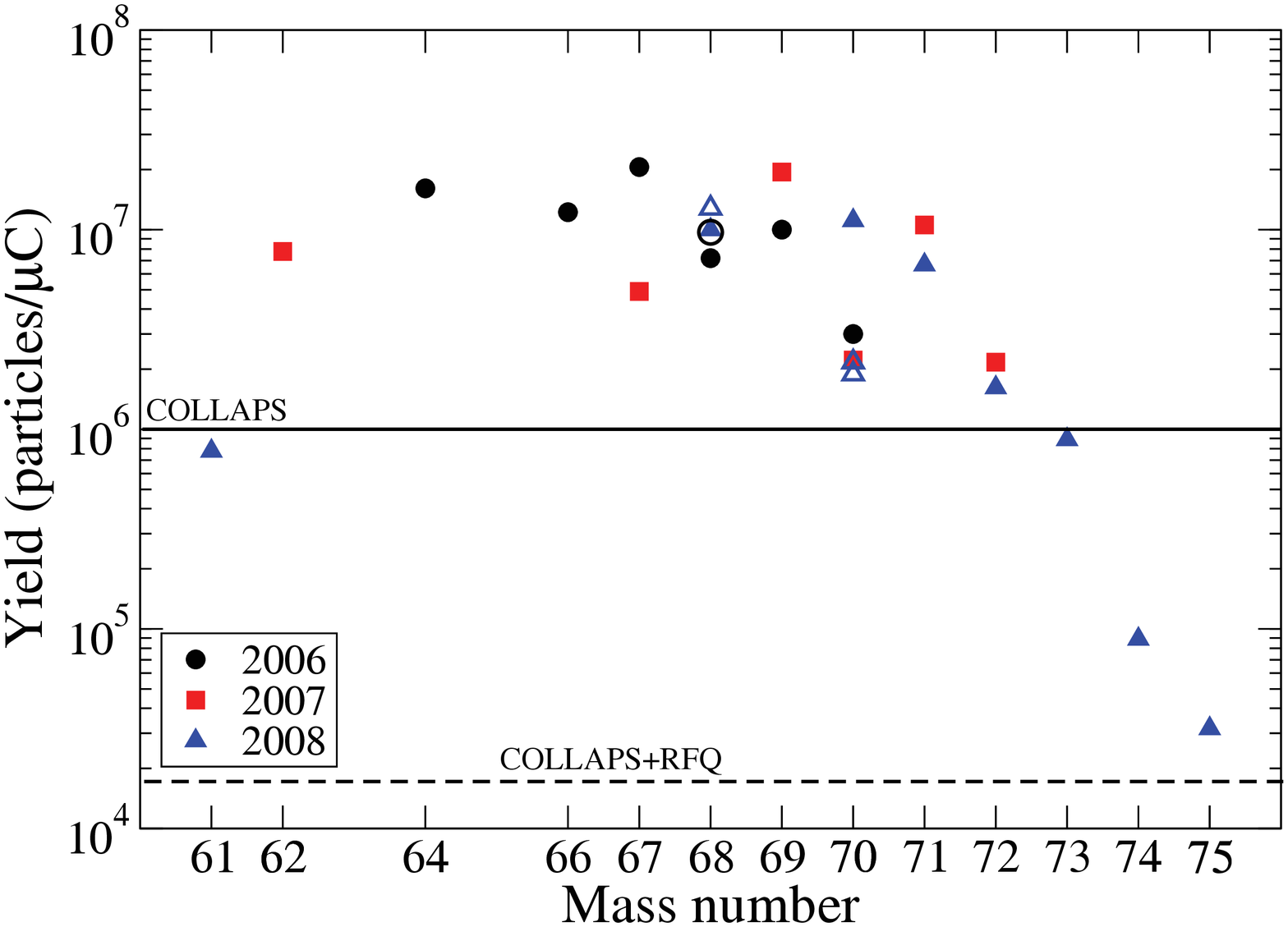}
\caption{\label{production}(Color online) Production yield of radioactive ground states (full symbols) and long lived isomers (open symbols) during the experiments using the GPS (circles and squares) and using HRS (triangles). The limit for laser spectroscopy measurements before installation of the ISCOOL device is indicated with a solid line, the current limit is shown by a dashed line.}
\end{figure}

\begin{figure}
\includegraphics[scale = 0.3]{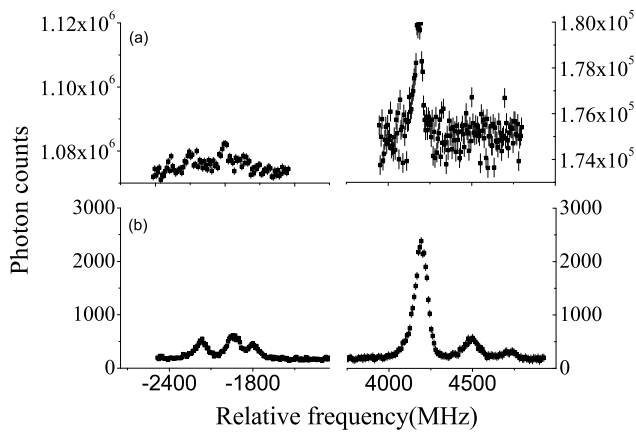}
\caption{\label{buncher}Fluorescence spectra for $^{72}$Cu. (a) Before installation of the cooler/buncher, after 10 hours of measurement. (b) With a bunched ion beam and photon gating, all 6 peaks are clearly resolved after 2 hours of measurement. }
\end{figure}

The experiment has been performed at the collinear laser spectroscopy setup COLLAPS at ISOLDE, CERN. Radioactive isotopes were produced by 1.4-GeV protons impinging on a 45 g/cm$^{2}$ thick uranium carbide target. The average proton beam current was 1.8 $\mu$A. Radioactive atoms were transported from the heated target through effusion and diffusion processes to a thin capillary tube, where they were step-wise ionized by the Resonance Ionization Laser Ion Source (RILIS). This was achieved in a two-step excitation scheme using the 327.4~nm \nolinebreak{$3d^{10}4s\,\,\,^2S_{1/2} \rightarrow 3d^{10}4p\,\,\, ^2P_{1/2}$} transition, followed by a 287.9 nm transition into an auto-ionizing state \cite{Kos00}. The ions were then accelerated and mass-separated, using either the general purpose isotope separator (GPS) or the high-resolution isotope separator (HRS). The calibration of the acceleration voltage is described in \cite{Kri10}. The copper ions were overlapped with the laser beam in the COLLAPS beam line by electrostatic deflectors (Fig.~\ref{setup}). The ions were neutralized in a sodium charge-exchange cell heated to approximately 220$^\circ$C. Subsequently, the atom beam could be resonantly excited from the $^{2}S_{1/2}$ atomic ground state to its $^2P_{3/2}$ state with a transition wavelength of 324.754 nm. The fluorescence was observed with two photomultiplier tubes (PMTs). Instead of scanning the laser frequency, a tunable post-acceleration voltage of $\pm$10kV was applied to the charge-exchange cell to obtain the resonance condition for the neutral copper atoms via Doppler tuning.
A summary of the experimental beamtimes is given in Table \ref{experiments}. Typical $^A$Cu beam intensities observed during the three experimental runs are shown in Fig.~\ref{production}. As for most exotic isotopes the beam current was too small to be recorded with the Faraday cup, the beam intensities were determined from the experimental efficiency as follows. For a stable $^{65}$Cu beam, the experimental efficiency was given by:
\begin{equation}
\epsilon_{\rm{COLLAPS}}=\frac{N_{\rm{COLLAPS}}}{N_{\rm{FC}}}
\end{equation}
with $N_{\rm{COLLAPS}}$ the amount of resonant photons per second observed in the strongest hyperfine component with the photomultiplier tubes, and $N_{\rm{FC}}$ the ions per second as observed by a Faraday cup placed after the mass separator. This experimental efficiency, which was typically 1:10000, was then used to calculate the production rate for the other isotopes given in Fig.~\ref{production}.
In the first two runs, the radioactive beam was accelerated to 50~keV and mass separated by the GPS. With a continuous ion beam, the optical detection of the resonant fluorescence was limited to ion beams of several  $10^6$~pps, due to stray light from the laser beam.  After installation of the linear gas-filled Paul trap ISCOOL \cite{Fra08, Man09} behind the HRS, measurements could be extended to more exotic isotopes with rates of a few $10^4$ pps. With ISCOOL the ions were trapped for typically 100 ms, and then released in a short pulse with a temporal width of $\sim$25 $\mu$s. By putting a gate on the PMT photon counting, accepting counts only when a bunch passed in front of the PMT, a reduction of the non-resonant photon background by a factor of 4000 is achieved.  This greatly improved the quality of the observed resonances, reduced the scanning time significantly and allowed investigation of more exotic isotopes. An example is given in Fig.~\ref{buncher} for measurements on $^{72}$Cu.

The laser system consisted of an Ar ion or Verdi pump laser and a Coherent 699 CW ring dye laser. Two methods have been used to correct for possible drifts in laser frequency during the experiments.  In experiments with a continuous beam and the GPS mass separator, a reference isotope was scanned during each scan of a radioactive isotope, by fast switching the mass selection of the GPS magnet. Because the mass change with the HRS magnets is not fast enough, during experiments with the ISCOOL buncher the frequency of the dye laser was locked to the iodine line at 15406.9373 cm$^{-1}$ using frequency modulation saturation spectroscopy. The exact frequency was measured with a Menlo systems frequency comb, and a frequency drift of less than 500~kHz (or 2$\cdot$10$^{-5}$ cm$^{-1}$) was observed during the experiment. A Spectra-Physics Wavetrain external cavity frequency doubler was used for second harmonic generation. A typical value of the laser power in the beamline was 1-2~mW.

\section{\label{result}Results}

Laser spectroscopy allows an accurate determination of magnetic moments and spectroscopic quadrupole moments, as well as isotopic changes in the mean square charge radius. In free atoms, the electronic levels are split with respect to their fine-structure energy by the hyperfine interaction, according to
\begin{equation}
\label{eqAB}
E_F = \frac{1}{2}AC+B\frac{\frac{3}{4}C(C+1)-I(I+1)J(J+1)}{2I(2I-1)J(2J-1)}
\end{equation}
with the quantum number $F$ given by \textbf{F}=\textbf{I}+\textbf{J}, $I$ being the nuclear and $J$ the electronic angular momentum, and
\begin{equation}
C = F(F+1)-I(I+1)-J(J+1).
\end{equation}
$A=\frac{\mu B_J}{IJ}$ depends on the nuclear magnetic moment $\mu$, with $B_J$ the magnetic field of the electrons at the position of the nucleus. $B=Q_s V_{zz}$ is related to the spectroscopic quadrupole moment $Q_s$, with $V_{zz}$ the electric field gradient created by the atomic electrons at the nucleus.

\begin{figure}[ht]
\includegraphics[scale = 0.38]{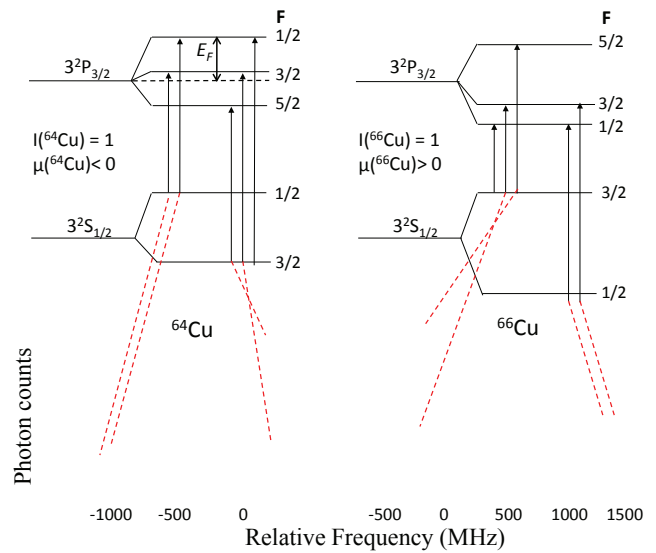}
\caption{\label{laserspec} (color online) Top: hyperfine splitting of the $^{2}S_{1/2}$ and $^{2}P_{3/2}$ levels for a nuclear spin $I=1$. The hyperfine energy $E_F$ is given relative to the fine-structure energy. Bottom: hyperfine spectra for $^{64}$Cu and $^{66}$Cu.  The sign of the magnetic and quadrupole moment is unambiguously determined by the positions and relative intensities of the resonances.}
\end{figure}

$V_{zz}$ is only non-zero in atomic states with $J>$1/2. In order to be able to extract the quadrupole moment, the $^{2}S_{1/2}\rightarrow^{2}P_{3/2}$ transition was therefore used in the measurements reported here. This transition is also sensitive to the nuclear spin. The sign of the nuclear moment is determined from the measured hyperfine spectrum, as shown in Fig.~\ref{laserspec} for $^{64}$Cu and $^{66}$Cu. In the case of a nuclear spin 1, five allowed transitions can be induced between the ground state and the excited hyperfine-split levels.

\begin{table*}[htb]
\caption{\label{tabmuQ1} Collinear laser spectroscopy results for copper ground and isomeric states. For the long-lived isomeric states the excitation energies are given \cite{Van04,Ste07}. The obtained values for $^{63,65}$Cu are in agreement with literature \cite{Tin57,Ney66}. For $^{62}$Cu, $^{64}$Cu and $^{66}$Cu the ratio of hyperfine parameters was fixed to the literature value 30.13(2)\cite{Tin57,Ney66}. The results of $^{71-75}$Cu have already been published in \cite{Fla09,Fla10}.}
\begin{ruledtabular}
\begin{tabular}{ccc| ccc}
Isotope &I$^{\pi}$ & Ex(keV)&  $A(^{2}S_{1/2})$(MHz) &$A(^2P_{3/2})$(MHz)  & $B(^{2}P_{3/2})$(MHz) \\
\hline
$^{61}$Cu & 3/2$^{-}$&0 & +5564(3)    & +185.5(10)       & -28(3)  \\
$^{62}$Cu & 1$^{+}$   &0& -1508(5)   &  FIXED      & -1(3)  \\
$^{63}$Cu & 3/2$^{-}$ &0& +5867.1(5)  &  +194.5(11)      & -28.0(6)  \\
$^{64}$Cu & 1$^{+}$   &0& -856.6(15)    & FIXED       & +9.6(12)  \\
$^{65}$Cu & 3/2$^{-}$ &0& +6284.0(7)  &  +208.4(2)      & -25.9(4)   \\
$^{66}$Cu & 1$^{+}$   &0& +1117(3)    &  FIXED      & +7(2)  \\
$^{67}$Cu & 3/2$^{-}$ &0& +6634.1(11) &   +220.2(5)     & -23.1(9)  \\
$^{68g}$Cu & 1$^{+}$  &0 & +9472.4(19)    &  +313.0(7)      &-11(2)  \\
$^{68m}$Cu & 6$^{-}$  &722 & +761.8(4)   &   +25.40(16)     &-59(2)  \\
$^{69}$Cu & 3/2$^{-}$ &0& +7489(2)    &   +248.7(15)     & -20(2)   \\
$^{70g}$Cu & 6$^{-}$   &0& +901.5(3)   &   +30.06(13)     & -37.8(14)  \\
$^{70m1}$Cu & 3$^{-}$   &101& -4438.1(18)&   -147.7(7)     & -18(6)  \\
$^{70m2}$Cu & 1$^{+}$   &242& +7037(6)    &   +234.4(17)     & -16(4)  \\
$^{71}$Cu & 3/2$^{-}$ &0& +6002(2)    &   +199.6(8)     & -25.3(14)  \\
$^{72}$Cu & 2$^{-}$   &0& -2666(2)   &   -89.8(6)     & +10(2)  \\
$^{73}$Cu & 3/2$^{-}$ &0& +4598(2)    &   +152.4(3)     & -26.5(10)  \\
$^{74}$Cu & 2$^{-}$   &0& -2113(5)   &   -71.6(11)     & +34(4)  \\
$^{75}$Cu & 5/2$^{-}$ &0& +1593(2)    &   +53.0(9)     & -36(2)  \\
\end{tabular}
\end{ruledtabular}
\end{table*}

Fitting of the spectra was done with Lorentzian functions assuming equal widths and with the peak intensities as free parameters.  The relative peak positions were constrained by equation (\ref{eqAB}).  The other fit parameters were the hyperfine parameters A($^2S_{1/2}$), A($^2P_{3/2}$), B($^2P_{3/2}$) and the center of gravity of the hyperfine structure. The fit was performed with an assumed value of the nuclear spin. A ROOT script employing the standard MINUIT fit package was used to fit the spectra. The fitting uncertainties on the parameters were multiplied by the square root of the reduced chi-squared of the fit. For most isotopes several independent measurements were performed, in which case the weighted average of the individual results was taken, and the error on the weighted average was taken as:
\begin{equation}
\sigma_{\rm{total}}=\max(\sigma_{\rm{fit}},\sigma/\sqrt{N})
\end{equation}
with $\sigma_{\rm{fit}}$ the statistical error on the weighted average due to the fitting uncertainties on the individual spectra, $\sigma$ the standard deviation of the weighted average and $N$ the number of independent scans. For almost all results, $\sigma_{\rm{fit}}$ was larger than $\sigma/\sqrt{N}$. A systematic error corresponding to an uncertainty on the acceleration voltage was taken into account, however it was significantly smaller than the statistical error. The results for all isotopes and isomers are listed in Table \ref{tabmuQ1}. The values for the stable isotopes $^{63,65}$Cu are in agreement with the literature values \cite{Tin57,Ney66}.

\begin{figure}
\includegraphics[scale=0.3]{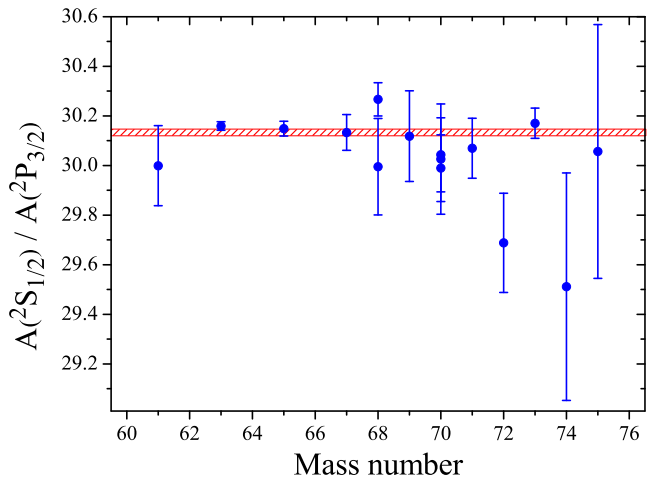}
\caption{\label{ratio} (color online) Ratio of the $A$-factors obtained from fitting the hyperfine spectra with their assigned nuclear spin.  All values are in agreement within 2$\sigma$ with the literature value of 30.13(2) (horizontal bar).}
\end{figure}

As the hyperfine anomaly between $^{63}$Cu and $^{65}$Cu was estimated to be less than 5$\cdot$10$^{-5}$~\cite{Loc74}, which is an order of magnitude less than our measurement accuracy, the ratio of the hyperfine parameters  $A(^{2}S_{1/2})$/$A(^{2}P_{3/2})$ is expected to be constant across the isotope chain, provided the correct nuclear spin has been assumed. This is illustrated in Fig.~\ref{ratio} for the nuclear spins given in Table \ref{tabmuQ1}. All experimental ratios are in agreement with the literature value of 30.13(2)~\cite{Tin57,Ney66}, which is indicated by the horizontal bar. Fitting with a different nuclear spin assumption led to a significant deviation from this average ratio, and allowed to firmly establish the ground-state spins of the exotic isotopes $^{71-75}$Cu \cite{Fla09,Fla10}. The deduced spins for the other isotopes are in agreement with the literature values \cite{NNDC}. For the odd-odd isotopes $^{62,64,66}$Cu, the statistics in the spectra did not allow a fit with a free ratio of $A$-parameters, and these spectra were fitted with the ratio fixed to 30.13.

From the $A$ and $B$ factors in Table II, we can determine the magnetic dipole and electric quadrupole moments of the isotopes and isomers relative to those of a reference isotope:
\begin{equation}
\label{eqmu}
\mu = \frac{AI}{A_{\rm ref}I_{\rm ref}}\mu_{\rm ref}
\end{equation}
\begin{equation}
\label{eqQ}
Q=\frac{B}{B_{\rm ref}}Q_{\rm ref}.
\end{equation}

\begin{center}
\begin{figure}[htb]
\includegraphics[scale=0.35]{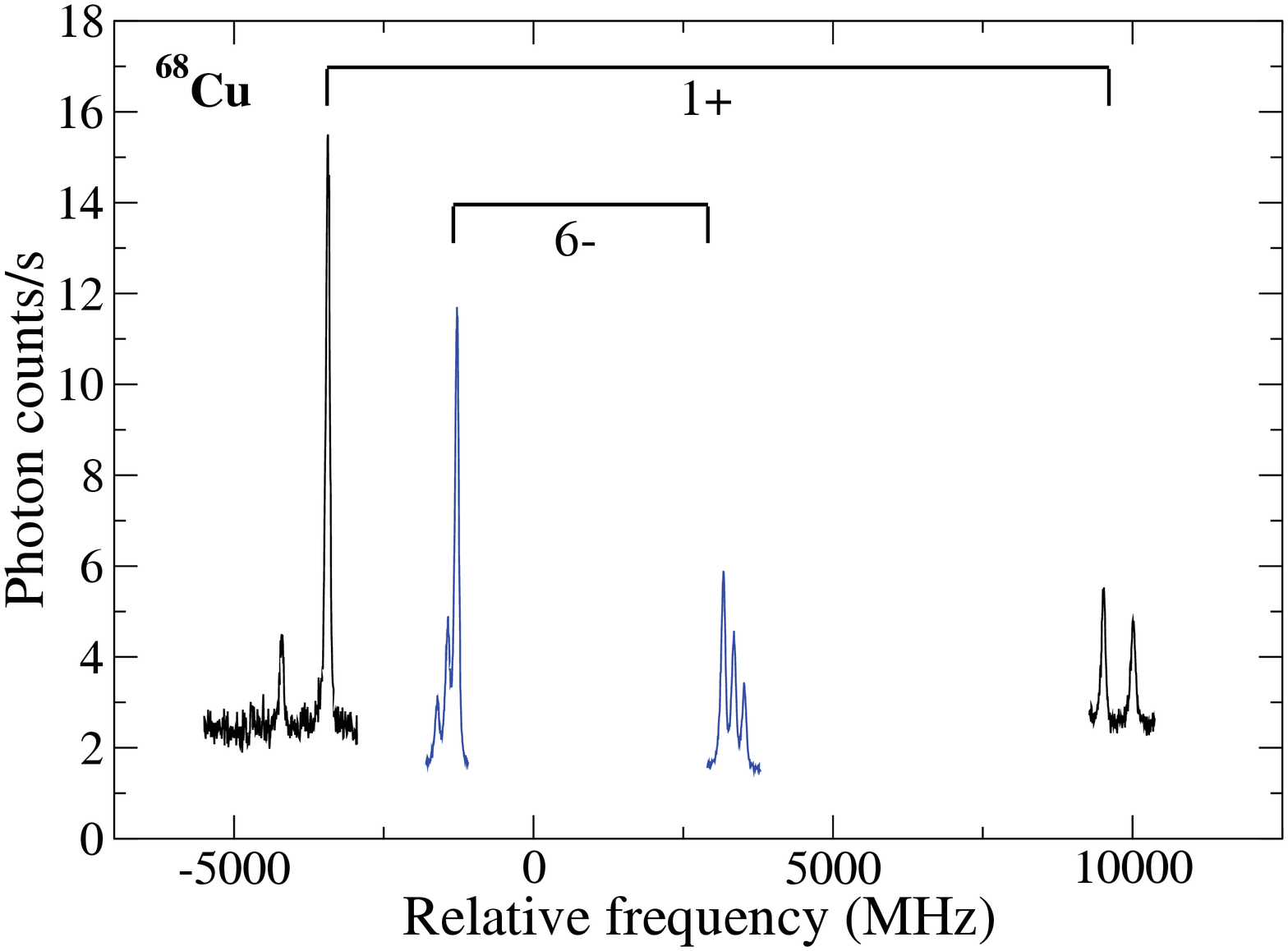}
\caption{\label{68Cu}(color online) Hyperfine spectrum for $^{68}$Cu. The transition lines for the ground and isomeric state are shown. The zero of the frequency scale corresponds to the center of gravity of the $^{65}$Cu hyperfine structure.}
\end{figure}
\end{center}

We used the literature values for the stable $^{65}$Cu isotope as a reference, with $A_{\rm ref}$=+6284.405(5)~MHz, $B_{\rm ref}$= --25.9(4)~MHz, \nolinebreak{$\mu_{\rm ref}$= +2.3817(3)~$\mu_{N}$} and $Q_{\rm ref}=-19.5(4)$~efm$^{2}$~\cite{Tin57,Ney66,Rag89}. The deduced magnetic dipole and electric quadrupole moments are shown in Table \ref{tabmuQ2} and compared to earlier results. All values presented here (except for $^{66}$Cu) are in agreement with literature, but in most cases the precision has been greatly improved. For $^{66}$Cu the sign of the magnetic moment was previously incorrectly assigned, and was found to be positive instead of negative \cite{Rag89}. As shown in Fig.~\ref{laserspec} for $^{64,66}$Cu, the sign of the magnetic moment determines the ordering of the hyperfine levels, which is observed directly through the intensity and position of the resonances. For $^{68}$Cu and $^{70}$Cu, low-lying isomers were observed, which were previously identified \cite{Wei02, Blau04, Van04}. Their moments were measured with in-source laser spectroscopy yielding values with very low precision because only the ground-state splitting was resolved \cite{Wei02,Ghe04}. The results of \cite{Ghe04} are shown in table \ref{tabmuQ2}, as the errors in \cite{Wei02} appear to be underestimated. With the technique of collinear laser spectroscopy all transitions can be resolved (Fig. \ref{68Cu}). In that way, the error on the magnetic moments is reduced by three orders of magnitude and the quadrupole moments can be obtained as well.

\begin{table*}[htb]
\caption{\label{tabmuQ2}Nuclear moments, deduced relative to the stable $^{65}$Cu, show an excellent agreement with the literature values.}
\begin{ruledtabular}
\begin{tabular}{cc| ccc | ccc}
Isotope & I$^{\pi}$  & $\mu_{exp}$($\mu_N$)& $\mu_{lit}$($\mu_N$)&ref & $Q_{exp}(\rm efm^{2}$)& $Q_{lit}(\rm efm^{2}$) &ref\\
\hline
$^{57}$Cu & 3/2$^{-}$  &  & +2.582(7) & \cite{Coc09} & & &\\
$^{59}$Cu & 3/2$^{-}$  &  & +1.910(4) &\cite{Coc09} & & &\\
          &           &   & +1.891(9) &\cite{Gol04} & & &\\
$^{61}$Cu & 3/2$^{-}$  & +2.1089(11) & +2.14(4) &\cite{Rag89} & -21(2)& &\\
$^{63}$Cu & 3/2$^{-}$   & +2.2236(4) & +2.22329(18)&\cite{Rag89} & -21.1(7) & -21.1(4)&\cite{Rag89}\\
$^{65}$Cu & 3/2$^{-}$  & & +2.38167(25)&\cite{Rag89} &  & -19.5(4)&\cite{Rag89}\\
$^{67}$Cu & 3/2$^{-}$   & +2.5142(6) & +2.54(2) &\cite{Rik00b} & -17.4(8) && \\
$^{69}$Cu & 3/2$^{-}$   & +2.8383(10) & +2.84(1)&\cite{Rik00} & -14.7(16)&&\\
$^{71}$Cu & 3/2$^{-}$   & +2.2747(8) & +2.28(1)&\cite{Sto08}& -19.0(16) &&\\
$^{73}$Cu & 3/2$^{-}$  & +1.7426(8) & &  & -20.0(10)&&\\
$^{75}$Cu & 5/2$^{-}$   & +1.0062(13) &&  & -26.9(16)&&\\
\hline
$^{58}$Cu & 1$^{+}$  &  & +0.479(13)&\cite{Coc10} & &&\\
$^{60}$Cu & 2$^{+}$  &  & +1.219(3)&\cite{Rag89} & &&\\
$^{62}$Cu & 1$^{+}$  & -0.3809(12) & -0.380(4) &\cite{Rag89} & 0(2)&&\\
$^{64}$Cu & 1$^{+}$   & -0.2164(4) & -0.217(2) &\cite{Rag89} & +7.2(9)&&\\
$^{66}$Cu & 1$^{+}$   & +0.2823(8) & -0.282(2) &\cite{Rag89}& +5.6(13) &&\\
$^{68g}$Cu & 1$^{+}$   & +2.3933(6) & +2.55(8)(19) &\cite{Ghe04} & -8.2(13) &&\\
$^{68m}$Cu & 6$^{-}$   & +1.1548(6) &  +1.26(7)(55) &\cite{Ghe04} & -44.0(19)   &&\\
$^{70g}$Cu & 6$^{-}$   & +1.3666(5) & +1.58(9)(57) &\cite{Ghe04} & -28.5(14)   &&\\
$^{70m1}$Cu & 3$^{-}$   & -3.3641(15) & -3.54(8)(34)&\cite{Ghe04} & -13(4)    &&\\
$^{70m2}$Cu & 1$^{+}$   & +1.7779(15) & +1.89(4)(14)&\cite{Ghe04} & -12(3)  &&\\
$^{72}$Cu & 2$^{-}$   & -1.3472(10) && & +8(2) &&\\
$^{74}$Cu & 2$^{-}$   & -1.068(3) & & & +26(3)&&\\
\end{tabular}
\end{ruledtabular}
\end{table*}

\section{\label{discussion}Discussion}

The experimental moments over the long chain of copper isotopes from $N=28$ to $N=46$ are a good testing ground for theoretical calculations as they span a broad range in neutron number. The most neutron-rich isotopes require the inclusion of the $\nu g_{9/2}$ orbital in the model space for shell-model calculations, increasing drastically the dimensionality of the problem.  For the isotopes up to $^{69}$Cu, it has been shown that the full $pf$ shell is sufficient to describe their ground-state magnetic moments \cite{Coc09}. That is because a single-particle excitation to the $\nu g_{9/2}$ orbital is forbidden due to the parity change.  Several effective shell-model interactions have been developed for the $pf$ shell (see~\cite{Hon02} for an overview). The most recent one, GXPF1 and its modifications \cite{Hon02,Hon04}, allows description of the properties of many isotopes in this region from $^{40}$Ca up to the heavier isotopes around $^{56}$Ni.

\begin{figure}[htb]
\includegraphics[scale=0.35]{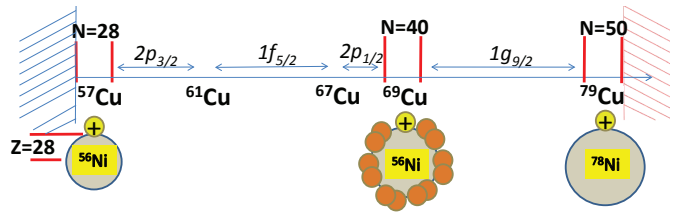}
\caption{(color online) The model space of the jj44b and JUN45 interactions consists of a $^{56}$Ni core, with copper isotopes having one proton outside the magic $Z=28$ shell. Beyond $^{57}$Cu$_{28}$ up to $^{79}$Cu$_{50}$, the negative parity neutron orbits $2p_{3/2}$, $1f_{5/2}$, $2p_{1/2}$ and the positive parity $1g_{9/2}$ are filled, from the $N=28$ to the $N=50$ shell gap across the $N=40$ harmonic oscillator sub-shell gap.}
\label{Curegion}
\end{figure}

Here, we will compare the moments of the full copper chain with calculations in an extended model space including the $g_{9/2}$ orbital.  Two effective shell-model interactions have been developed for the $f5pg9$ model space starting from a $^{56}$Ni core (Fig.~\ref{Curegion}), excluding excitations of protons and neutrons across $N= Z=28$. The jj44b interaction is determined by fitting single particle energies and two-body matrix elements to data from the nickel and copper chain and data along the $N=50$ isotones \cite{Ver07}.  This interaction should thus give rather good results for the copper isotopes presented here. The JUN45 interaction on the other hand, has been fitted to experimental data of 69 nuclei with masses $A$=63 to $A$=96 in the upper $pf$ shell, excluding all nickel and copper isotopes \cite{Hon09}.  In the latter study its aim was to investigate the effect of the missing $f_{7/2}$ orbital on the calculated nuclear properties in this region.  Comparing the calculated magnetic and quadrupole moments, as well as the low-level structure of the copper isotopes with our experimental data will provide a good test for this study.
 A recent calculation with an effective interaction in an extended $pfg_{9/2}$ model space, starting from a $^{48}$Ca core \cite{Sie10}, very well reproduced the experimental magnetic moments of the odd-A Cu isotopes beyond $N=40$, while the two above models (jj44b, JUN45) based on the $^{56}$Ni core overestimated the $^{71,73}$Cu magnetic moments ~\cite{Fla09}.  Also, the observed steep lowering of the 1/2$^-$ level towards $^{75}$Cu was reproduced in \cite{Sie10}, while not in \cite{Fla09}.  This indeed suggests the need for including proton excitations across $Z=28$ in order to describe correctly the properties of the neutron-rich Cu isotopes.

In the next section we will discuss the properties of the odd-A Cu isotopes, whose ground-state moments are dominated by the odd proton occupying either the $\pi p_{3/2}$ or $\pi f_{5/2}$ level. The second part of the discussion will focus on the properties of the odd-odd Cu isotopes, which are entirely dependent on the coupling of the single proton to neutrons in the $f5pg9$ space.

\subsubsection{\label{sectionoddgfactors}The odd-A Cu isotopes}

The nuclear $g$-factor is a dimensionless quantity related to the magnetic moment via the nuclear spin:
\begin{equation}
\label{gfactor}
g = \frac{\mu}{I\mu_N}
\end{equation}
with $\mu_N$ the nuclear magneton. The $g$-factor is very sensitive to the orbital occupation of the unpaired nucleons \cite{Ney03}.

\begin{figure}[htb]
\includegraphics[scale=0.33]{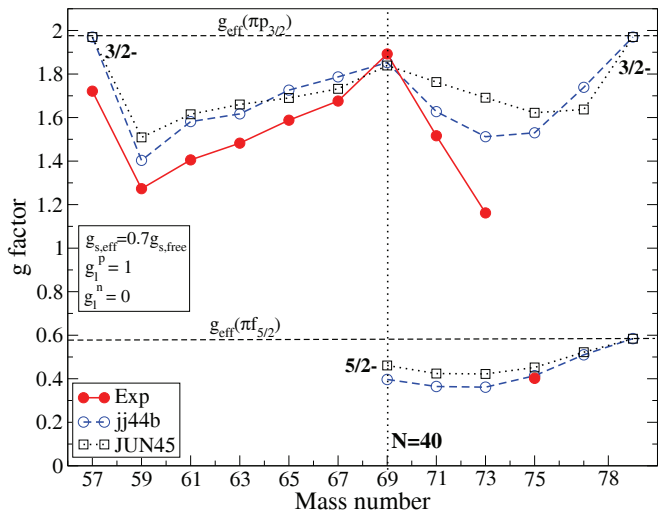}
\caption{ (color online) Experimental $g$-factors (filled dots) compared with calculations (open symbols) using the jj44b and JUN45 interactions \cite{Hon09}. An effective spin $g$-factor of 0.7$g_{\rm s,free}$ was adopted.}
\label{oddgfactorsc}
\end{figure}

\begin{figure*}[htb]
\subfigure{\includegraphics[scale=0.6]{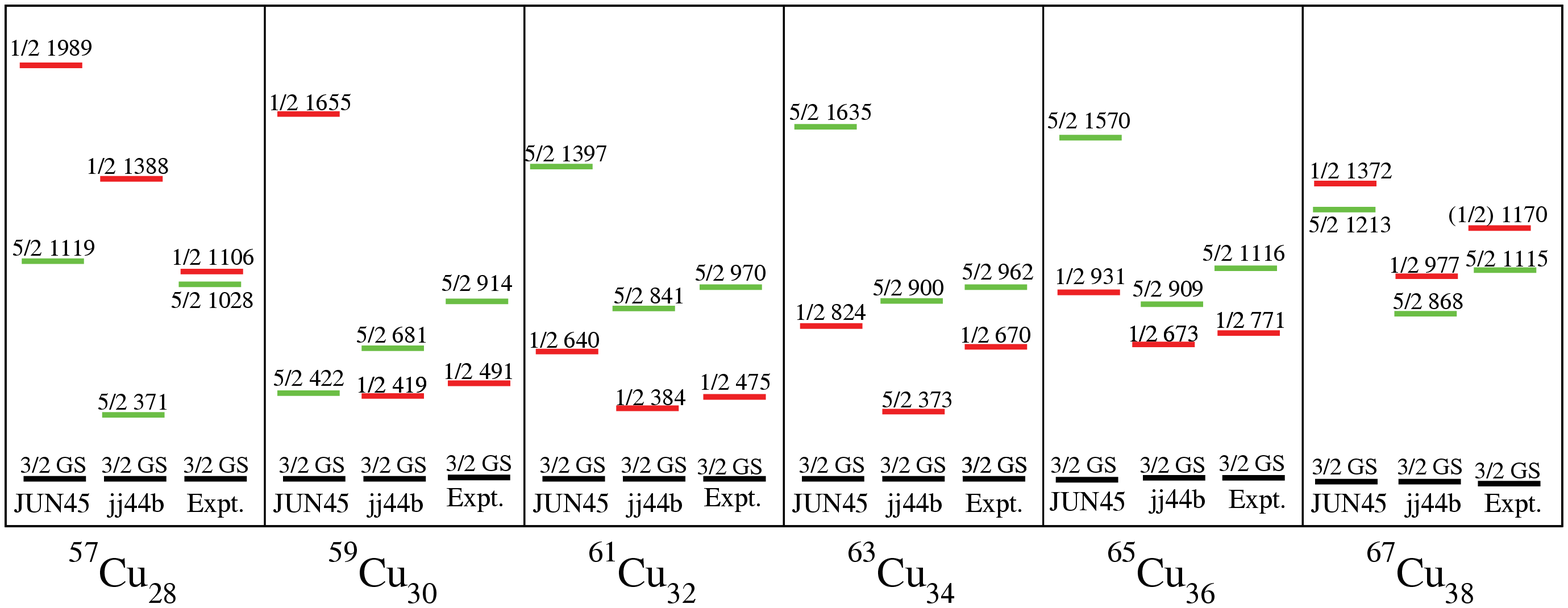}}
\subfigure{\includegraphics[scale=0.6]{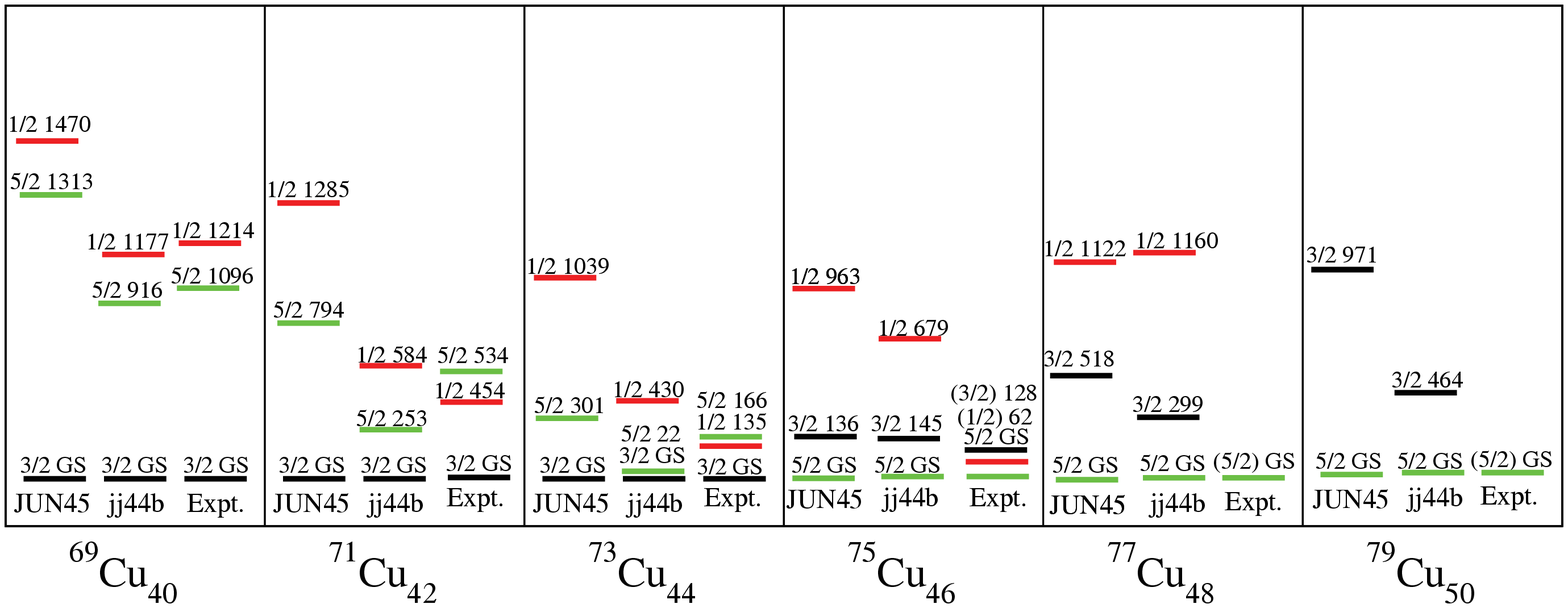}}
\caption{\label{figoddenergylevels}(color online) Experimental and calculated energy levels of odd-A Cu isotopes \cite{NNDC,Fra01,Ste08,Ste09,Dau10}. Only the lowest 1/2$^-$, 3/2$^-$ and 5/2$^-$ states are shown. }
\end{figure*}

In Fig.~\ref{oddgfactorsc} the odd-A Cu $g$-factors are compared with the results from both interactions. The experimental $g$-factors for $^{57}$Cu ($N=28$) and $^{69}$Cu ($N=40$) are closest to the effective single particle value for a $\pi p_{3/2}$ configuration. This suggests magicity at $N=28$ and $N=40$.  Indeed, in magic nuclei the ground state wave function can be approximated by a pure single particle configuration and so its $g$-factor will correspond to that of the pure configuration. The fact that $N=40$ appears here as a magic number, with the properties of a shell closure, is not only related to the magnitude of the energy gap between the $\nu p_{1/2}$ and the $\nu g_{9/2}$ single particle levels.  The magic behavior is mostly due to the parity change, which does not allow M1 excitations from the negative parity $pf$ orbits into the positive parity $\nu g_{9/2}$ orbital.

The fact that a rather large reduction of the effective single particle value is needed to reproduce the experimental $g$-factors is because small contributions of M1-excitations of the type $(f_{7/2}^{-1}f_{5/2})_{1^+}$ have a strong impact on the experimental $g$-factor. The fact that calculations without such excitations do not reproduce the experimental values towards $N=28$ (as in Fig.~\ref{oddgfactorsc}) is an indication that the $N=28$ gap is not very large.  This was already known from earlier studies \cite{Ken01,Hon04}.  Indeed, the experimental $g$-factors of the neutron-deficient Cu isotopes are very well reproduced with the GXPF1 shell model interaction in the full $pf$ space \cite{Hon04,Coc09}, which include proton and neutron excitations across $N=28$. Note that in this $pf$ model space, only a minor reduction of the $g_s$ factor is needed (0.9$g_{\rm s,free}$) to reproduce the experimental moments.

On the neutron-rich side of $N=40$, there is an unusually large discrepancy between theory and experiment for $^{73}$Cu, but not for $^{75}$Cu. This is probably because excitations of protons from the $\pi 1f_{7/2}$ level become increasingly important from $N=40$ onwards, as the gap between the $\pi 1f_{5/2}$ and the $\pi 1f_{7/2}$ levels decreases under the influence of the tensor force when the 1$g_{9/2}$ neutron orbit gets filled~\cite{Ots05,Sie10}. The $g$-factor of $^{75}$Cu is then well reproduced without including such proton excitations, because the $I=5/2$ ground state of $^{75}$Cu is dominated by a single proton in the 1$f_{5/2}$ orbit, thus blocking the ($\pi f_{7/2}^{-1} f_{5/2})_{1^+}$ mixing into the wave function.

The experimental energy levels for the odd-A Cu isotopes are compared with jj44b and JUN45 interactions in Fig.~\ref{figoddenergylevels}. The 3/2$^-$ state is the ground state for most isotopes, dominated by a proton in the $\pi 2p_{3/2}$ level. The spin inversion of the ground state from 3/2$^-$ to 5/2$^-$ at $^{75}$Cu, dominated by a proton in the $\pi 1f_{5/2}$ level \cite{Fla09}, is correctly reproduced by both interactions. The lowering of the $1/2^-$ level however is not so well reproduced, which might be attributed to missing proton excitations from the $\pi f_{7/2}$ level, as illustrated in \cite{Sie10}.  Towards $^{79}$Cu, the calculated energy spacing between the $5/2^-$ gs and the $3/2^-$ first excited state increases. At $N=50$, where the neutron space is completely filled, this gap is very sensitive to the effective single particle energy between the $\pi 1f_{5/2}$ and $\pi 2p_{3/2}$ orbits. Unfortunately no data are available for $^{79}$Cu, but in the $^{81}$Ga isotone this $(5/2^--3/2^-)$ energy spacing is measured to be 351 keV.  Both interactions reproduce this fairly well: 239 and 450 keV for jj44b and JUN45, respectively \cite{Che10}. Thus we can expect the experimental energy of the $3/2^-$ level in $^{79}$Cu between 464 keV and 971 keV. On the other hand, the recent shell-model calculation that takes into account proton excitations across $Z=28$ \cite{Sie10} predicts this $3/2^-$ level above 1.5 MeV \cite{Dau10}. Clearly, experimental data on the neutron-rich copper level schemes are required to further investigate this.

On the neutron deficient side, the jj44b interaction strongly underestimates the energy of the lowest 5/2$^-$ state in $^{57}$Cu, but all other $5/2^-$ levels are calculated within 200-300 keV from the experimental ones.  The $1/2^-$ level is fairly well reproduced up to $^{71}$Cu, but agreement diminishes for $^{73,75}$Cu.  The JUN45 interaction on the other hand, overestimates the energy of the 1/2$^-$ level in $^{57}$Cu by almost 1 MeV, while the agreement is better for the copper isotopes up to $^{69}$Cu. However, from $^{71}$Cu onwards, the 1/2$^-$ level is calculated again about 800 keV too high.

One of the questions raised in this region of the nuclear chart, is related to the onset of collectivity beyond $N=40$. The spectroscopic quadrupole moment is an ideal parameter to probe collectivity and reveal if the shape of nuclei strongly evolves when the number of available neutron correlations increases towards mid-shell, between $N=28$ and $N=50$.  As the jj44b and JUN45 interactions assume inert $^{56}$Ni and $^{78}$Ni cores, proton and neutron effective charges have to be used that take into account these limitations in the model space.  Experimental quadrupole moments were used to fit the effective proton and neutron charges (e$_{\pi}$ and e$_{\nu}$), considering that the spectroscopic quadrupole moment is given by:

\begin{equation}
\label{effectivecharges}
Q_s=e_{\pi}Q_p + e_{\nu}Q_n
\end{equation}
where $Q_p$ and $Q_n$ are the contributions to the calculated spectroscopic quadrupole moments from protons and neutrons, respectively.  For the JUN45 interaction, all known quadrupole moments in the model space have been taken into account and the best result was obtained for e$_{\pi}$=+1.5e, e$_{\nu}$= +1.1e \cite{Hon09}. The same effective charges have been adopted here to calculate quadrupole moments with the jj44b interaction, as the two interactions are active in the same model space.

For both calculations a harmonic oscillator potential was used with a mass dependent energy defined as:
\begin{equation}
\hbar\omega=41A^{-1/3}.
\end{equation}
This dependence was also used in \cite{Hon09} to fit the effective proton and neutron charges. If the alternative formula for the oscillator energy is used:
\begin{equation}\label{oscillatorB}
\hbar\omega=45A^{-1/3}-25A^{-2/3}
\end{equation}
the calculated quadrupole moments are about 5\% larger, and thus the fitted effective charges would be slightly lower. Note that in \cite{Che10}, the calculations with JUN45 and jj44b have both been performed using (\ref{oscillatorB}) for the oscillator parameter.

\begin{figure}[htb]
\includegraphics[scale=0.35]{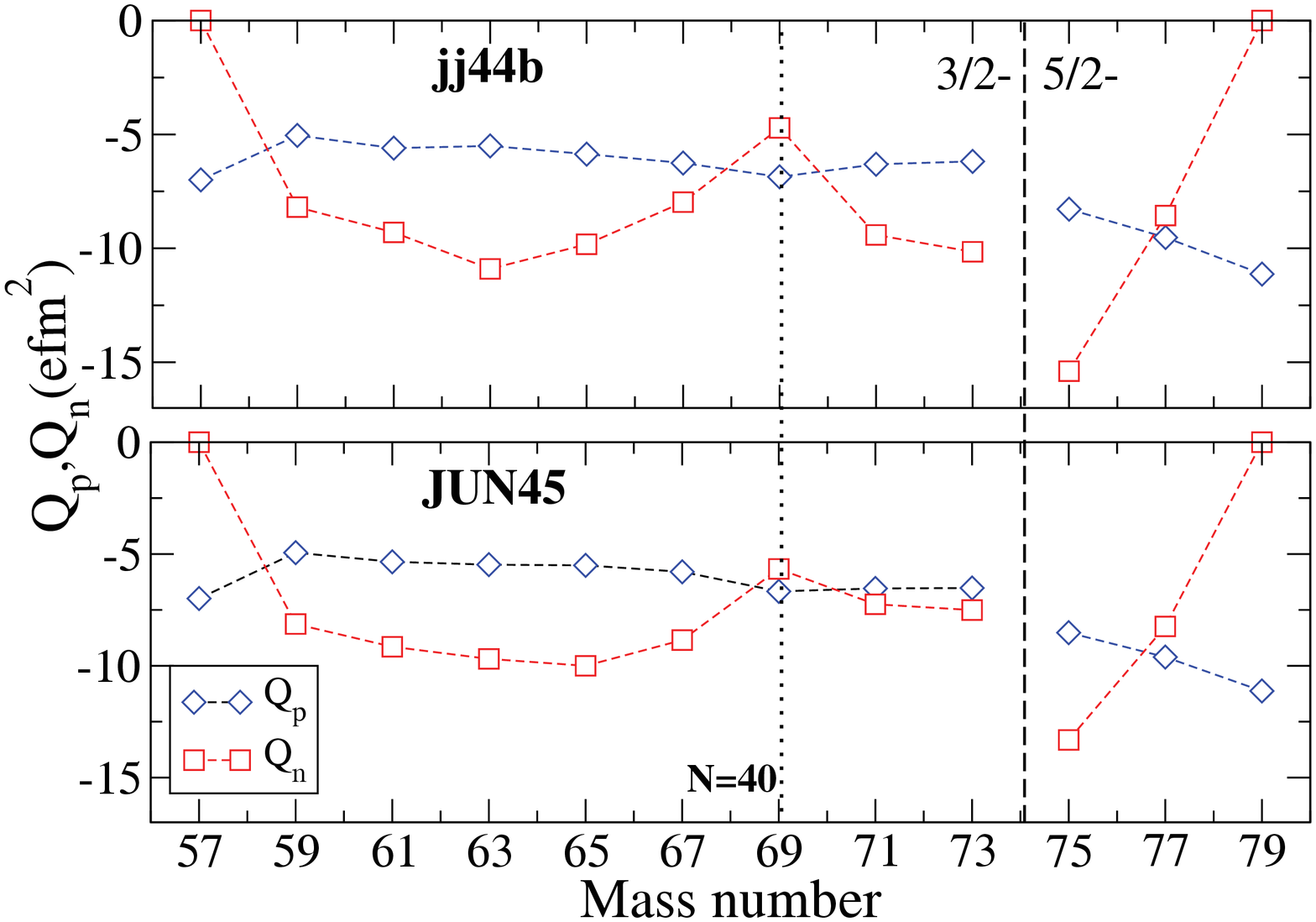}
\caption{\label{QpQn} (color online) The contributions to the spectroscopic quadrupole moment due to protons and neutrons separately, as given in Table \ref{tabQ}. The neutron contribution is responsible for the observed core polarizing effect when moving away from $N=40$.}
\end{figure}

\begin{table*}[htb]
\caption{\label{tabQ}Experimental and calculated quadrupole moments. The proton and neutron contributions to the theoretical quadrupole moment ($Q_p$ and $Q_n$) are given separately. $Q_{theo}$ is obtained with effective charges e$_{\pi}$ = +1.5e, e$_{\nu}$ = +1.1e.}
\begin{ruledtabular}
\begin{tabular}{|cc|c|ccc|ccc|}
&&\textbf{Exp}&&\textbf{jj44b}&&&\textbf{JUN45}& \\
Isotope & I$^{\pi}$  & $Q_{exp}(\rm efm^{2}$) & $Q_{p}(\rm efm^{2}$) & $Q_{n}(\rm efm^{2}$)& $Q_{theo}(\rm efm^{2}$)& $Q_{p}(\rm efm^{2}$) & $Q_{n}(\rm efm^{2}$)& $Q_{theo}(\rm efm^{2}$)\\
\hline
$^{57}$Cu & 3/2$^{-}$  & & -6.99&0&-10.49&-6.99& 0 &-10.49\\
$^{59}$Cu & 3/2$^{-}$  & & -5.04&-8.19&-16.56&-4.94&-8.13&-16.35\\
$^{61}$Cu & 3/2$^{-}$  & -21(2)& -5.60&-9.30&-18.63&-5.35&-9.15&-18.09\\
$^{63}$Cu & 3/2$^{-}$  & -21.1(4) & -5.50&-10.90&-20.24&-5.48&-9.69&-18.88\\
$^{65}$Cu & 3/2$^{-}$  & -19.5(4) & -5.86&-9.82&-19.59&-5.51&-10.00&-19.27\\
$^{67}$Cu & 3/2$^{-}$  & -17.4(8) & -6.25&-7.97&-18.15&-5.80&-8.85&-18.44\\
$^{69}$Cu & 3/2$^{-}$  & -14.7(16)& -6.86&-4.71&-15.47&-6.67&-5.67&-16.24\\
$^{71}$Cu & 3/2$^{-}$  & -19.0(16)& -6.31&-9.41&-19.81&-6.54&-7.25&-17.79 \\
$^{73}$Cu & 3/2$^{-}$  & -20.0(10)& -6.18&-10.15&-20.44&-6.52&-7.51&-18.04\\
$^{75}$Cu & 5/2$^{-}$  & -26.9(16)& -8.28&-15.39&-29.35&-8.52&-13.32&-27.43\\
$^{77}$Cu & 5/2$^{-}$  & & -9.52&-9.00&-23.70&-8.56&-8.24&-23.48\\
$^{79}$Cu & 5/2$^{-}$  & & -11.13&0&-16.70&-11.13& 0 &-16.70\\
\end{tabular}
\end{ruledtabular}
\end{table*}

\begin{figure}[htb]
\includegraphics[scale=0.35]{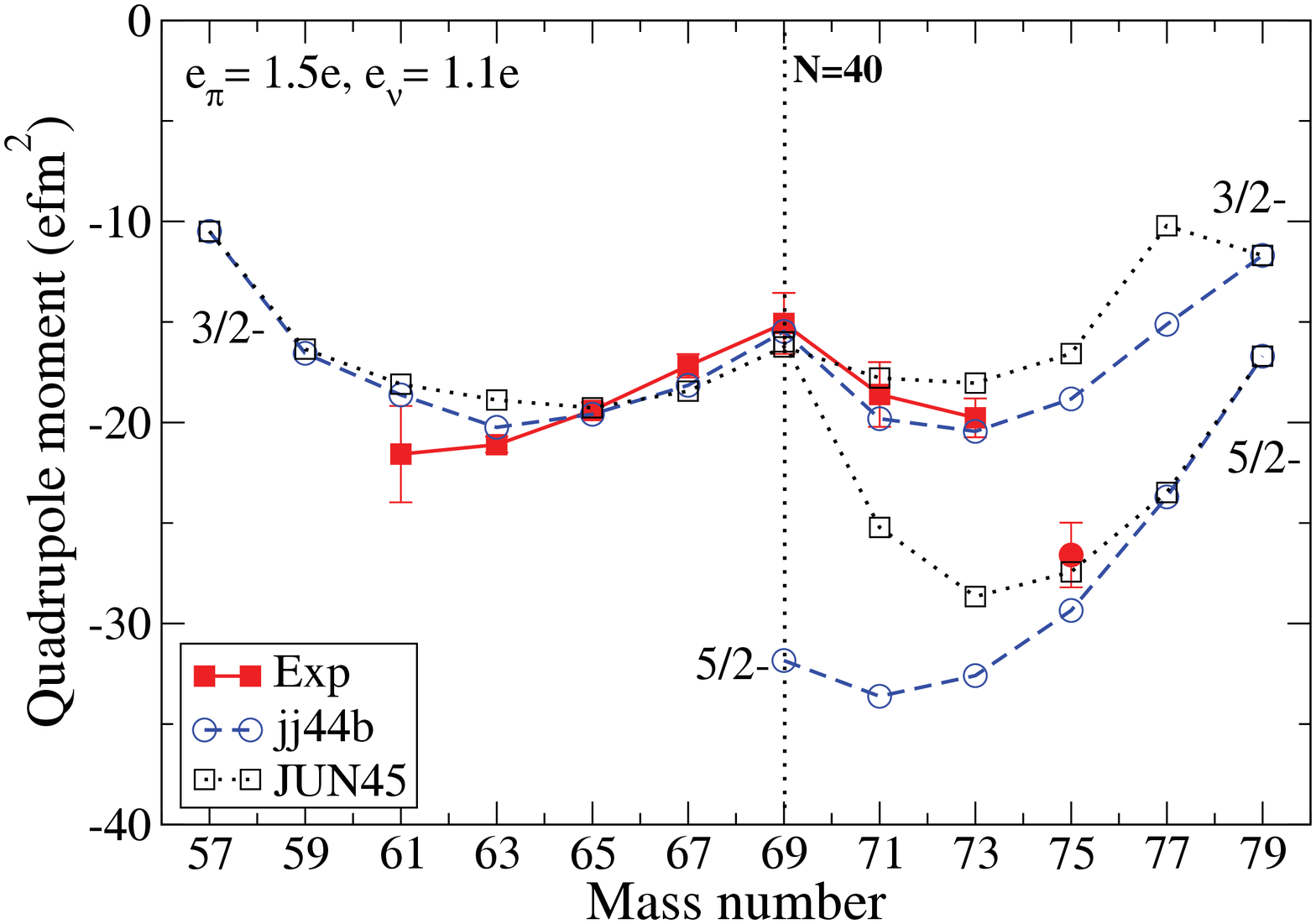}
\caption{(color online) The measured spectroscopic quadrupole moments compared with shell-model calculations. }
\label{figoddquadrupole1}
\end{figure}

In Fig.~\ref{QpQn} the calculated proton and neutron contributions ($Q_p$ and $Q_n$) to the spectroscopic quadrupole moments are shown for the two interactions (values given in Table \ref{tabQ}).
From Fig. \ref{QpQn} it is seen that the proton quadrupole moments of configurations dominated by a $\pi p_{3/2}$ proton are rather constant across the chain, which illustrates that in this case the proton-neutron correlation does not strongly affect $Q_p$.  In the neutron-rich isotopes, dominated by a $\pi f_{5/2}$ ground-state configuration, the proton quadrupole moment is more sensitive to this proton-neutron interaction. The trend is the same for both interactions, though the absolute proton quadrupole moments are slightly larger with jj44b. For the neutron quadrupole moments, a significant difference between the two calculations is seen in the region of the $N=40$ sub-shell gap. While both calculations show a strong core polarization when adding neutrons to $N=28$ or removing them from $N=50$, the reduction at $N=40$ is more pronounced for the jj44b interaction. The strong increase in core polarization beyond $N=40$, observed for the jj44b interaction, is missing for the JUN45 interaction. This is an indication that the $N=40$ gap is too small in the JUN45 effective interaction.

\begin{figure}[htb]
\subfigure{\includegraphics[scale=0.3]{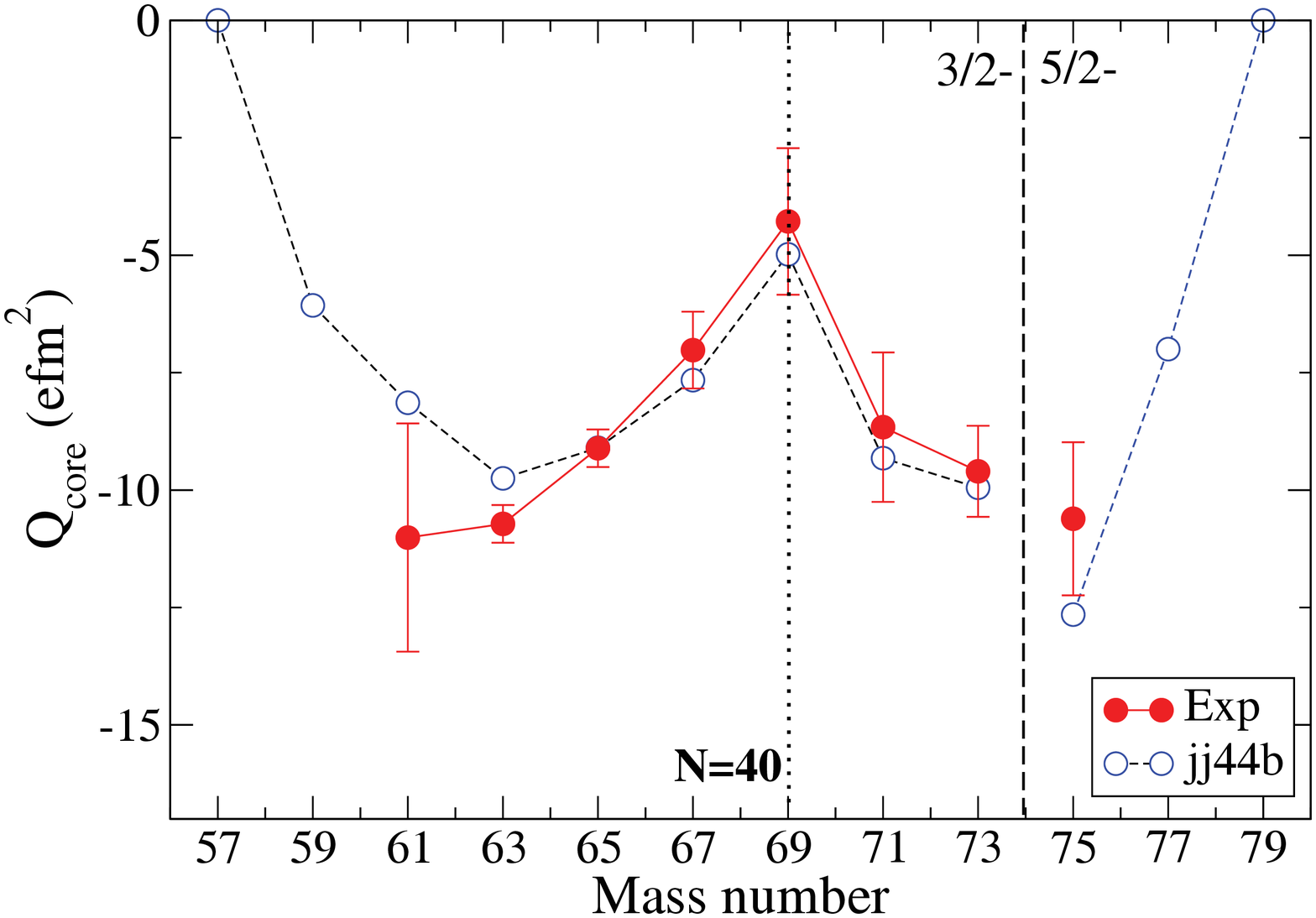}}
\subfigure{\includegraphics[scale=0.3]{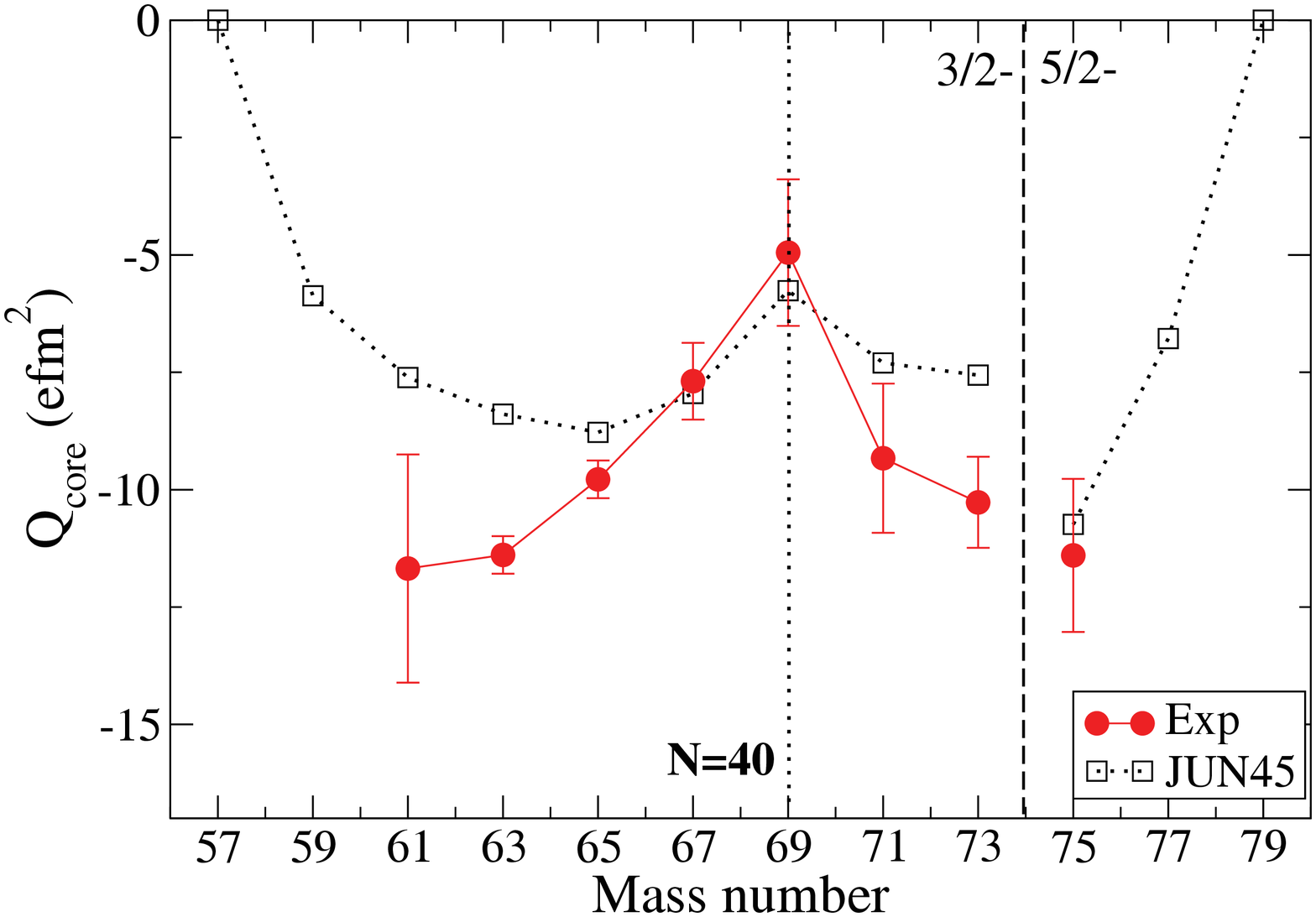}}
\caption{\label{Qcore}(color online) Experimental core polarization quadrupole moments compared to calculated core polarization moments with jj44b (top) and JUN45 (bottom). See text for details.}
\end{figure}

In Fig.~\ref{figoddquadrupole1} the experimental values are compared to calculated spectroscopic quadrupole moments.  A reasonable agreement is observed for both calculations.  The trend is however better reproduced by the jj44b interaction, which is clearly related to the stronger core-polarizing effect in the neutron quadrupole moments. When neutrons are added or removed from $N=40$, the experimental moments reveal a strong core-polarization effect. This core polarization is similar on both sides of $N=40$: the $^{65,67}$Cu and $^{71,73}$Cu spectroscopic quadrupole moments are the same within error bars. No increased collectivity is observed on the neutron-rich side, as suggested by recent measurements of the $B$(E2) transition rates in the underlying nickel isotopes \cite{Per06,Aoi10}.  In order to compare the core polarization in $^{75}$Cu to that of the other copper isotopes, we calculate the core quadrupole moment. This is done by taking the difference between the spectroscopic (experimental or calculated) value ($Q_s$) and the single particle moment ($Q_{\rm s.p.}$) for the odd proton:

\begin{equation}
\label{Qexp}
Q_{\rm core} = Q_{\rm s} - Q_{\rm s.p.}.
\end{equation}

The single particle quadrupole moment ($Q_{\rm s.p.}$) for the $3/2^-$ states is taken as the calculated effective moment for $^{57}$Cu, while the $5/2^-$ single particle moment is the calculated effective value for $^{79}$Cu.  Indeed, these isotopes have no valence neutrons and thus $Q_{\rm s.p.} = e_{\pi} Q_p$ (see Table \ref{tabQ} for values from jj44b and JUN45). The deduced experimental and calculated core polarizations are shown in Fig.~\ref{Qcore}. The experimental core polarization in $^{75}$Cu is the same as in $^{63}$Cu, so again no enhancement of collectivity is observed towards the neutron rich isotopes.  The jj44b interaction reproduces very well the trend in the core polarization. It seems to slightly overestimate the core polarization in $^{75}$Cu, but data on the more neutron-rich isotopes are needed to confirm this. Towards the neutron-deficient side, the core polarization seems underestimated in $^{61}$Cu.  However, a more precise experimental value and more precise determination of the effective proton charge is needed to establish this firmly.  For the JUN45 interaction, the core polarization is largely underestimated in all isotopes away from $^{67,69}$Cu, both when adding and when removing neutrons from $N=40$. Adjusting the effective charges does not improve the agreement with experiment.

\subsubsection{\label{evengfactors}The even-A Cu isotopes}

The structure of the even-A copper isotopes is dominated by the coupling between the odd proton in one of the $\pi 2p_{3/2}1f_{5/2}2p_{1/2}$ orbits and an unpaired neutron in one of the available neutron orbits.  This is illustrated well by comparing the experimental $g$-factors for the 1$^+$, 2$^+$ and $2^-$ states in $^{58-74}$Cu with empirical values calculated with the additivity relation for moments \cite{Ney03}, using experimental $g$-factors of neighboring even-odd nickel and zinc, and odd-even copper isotopes:

\begin{equation}
\label{empirical}
g(I) = \frac{g_{p}+g_{n}}{2}+\frac{(g_{p}-g_{n})}{2}\frac{j_p(j_p+1)-j_n(j_n+1)}{I(I+1)}.
\end{equation}

\begin{figure}
\includegraphics[scale=0.35]{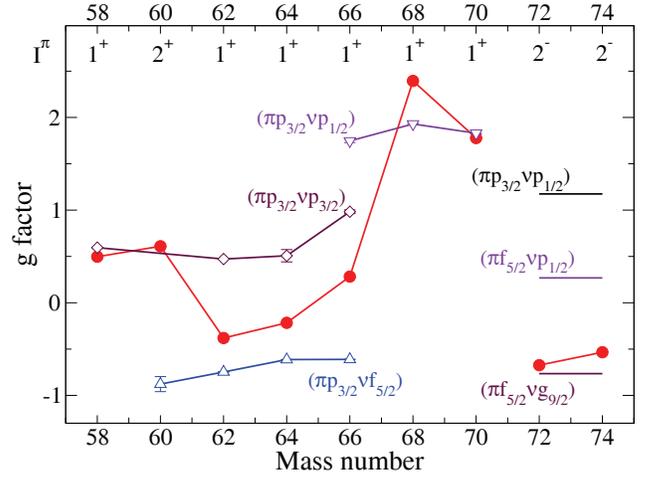}
\caption{(color online) The experimental $g$-factors of the ground states of $^{58-74}$Cu (Table \ref{tabmuQ2}, solid circles) are compared with empirical $g$-factors using the additivity relation for simple proton-neutron configurations. The positive sign of the $^{66}$Cu value reveals a strongly mixed ground-state wave function. }
\label{empiricalgfactors}
\end{figure}

\begin{figure*}[htb]
\subfigure{\includegraphics[scale=0.6]{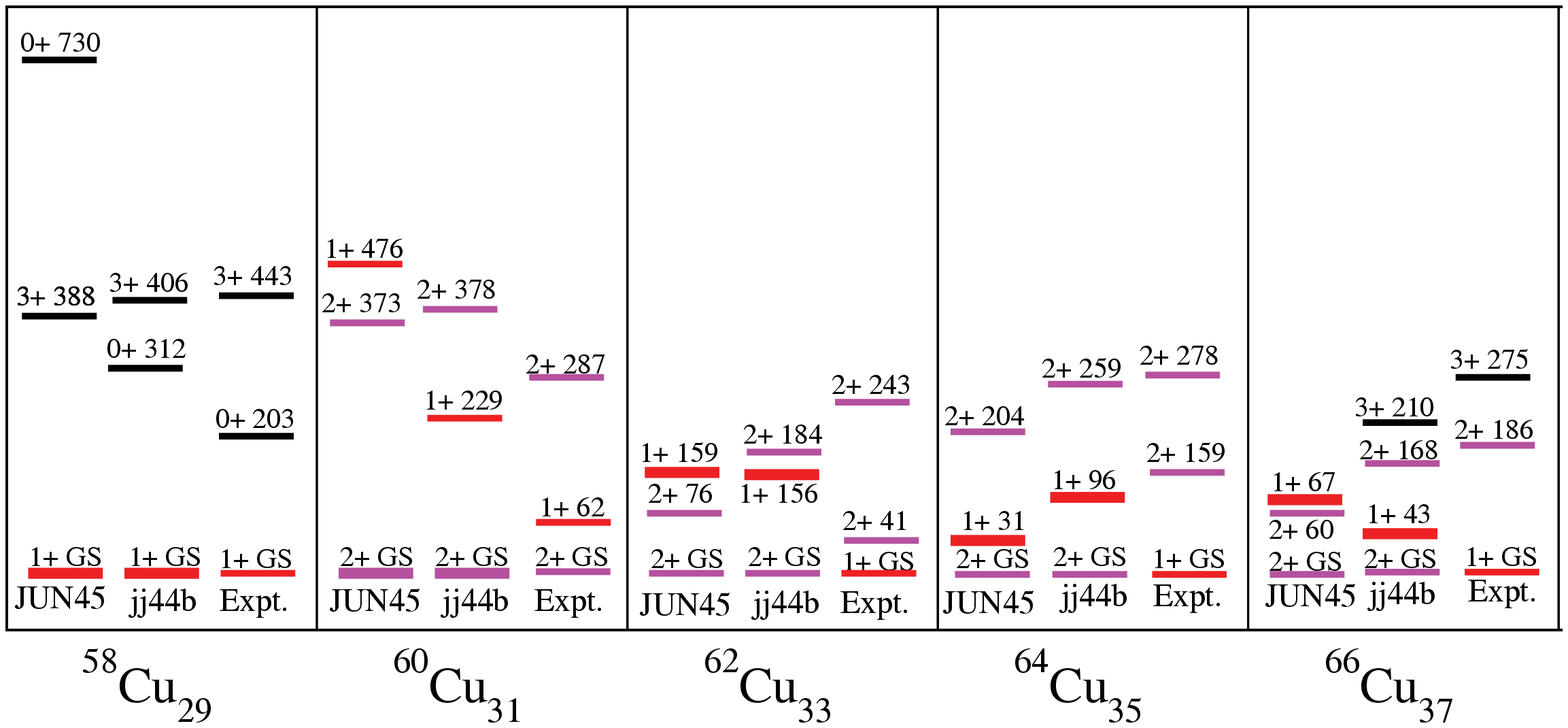}}
\subfigure{\includegraphics[scale=0.6]{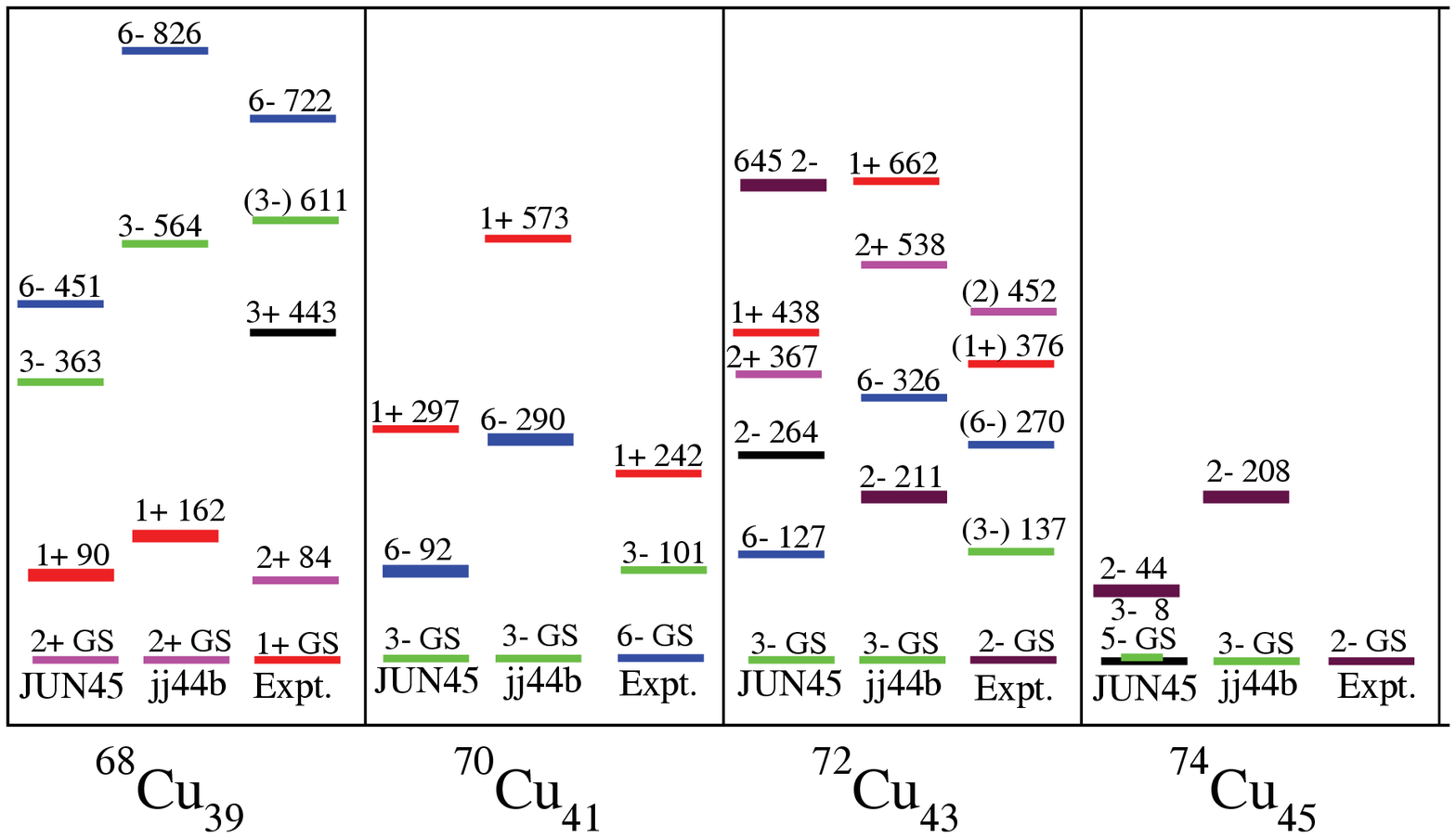}}
\caption{(color online) Experimental and calculated lowest levels in the odd-odd Cu isotopes \cite{NNDC, Van04, JCT06, Ste07}. Levels in bold have calculated moments in agreement with the experimentally measured magnetic and quadrupole moments of the ground state.}
\label{evenCulevels}
\end{figure*}

For example, the possible empirical moments for the 1$^+$ state in $^{66}$Cu for a $\pi p_{3/2}$ coupling to a $\nu p_{1/2}$, $\nu f_{5/2}$ or $\nu p_{3/2}$, are calculated using the experimental $g_n$ factors of the 1/2, 3/2 and 5/2 states in Zn \cite{Rag89} and for $g_p$ the average was taken between the $g$-factors of $^{65}$Cu and $^{67}$Cu. As shown in Fig.~\ref{empiricalgfactors}, the comparison of these empirical $g$-factors with the experimental values provides an indication for the purity of the state. $^{58}$Cu and $^{60}$Cu are clearly dominated by a $(\pi p_{3/2}\otimes\nu p_{3/2})$ configuration, while $^{62}$Cu and $^{64}$Cu have a dominant $(\pi p_{3/2}\otimes\nu f_{5/2})$ structure, as already concluded in \cite{Coc10}. The sign of the $^{66}$Cu $g$-factor was determined to be positive, in contrast to the literature value (see Table \ref{tabmuQ2}). This illustrates that $^{66}$Cu has a strongly mixed ground state, with a significant occupation of the $\nu p_{1/2}$ orbital.  The $1^+$ states in $^{68}$Cu and $^{70}$Cu show a $(\pi p_{3/2}\otimes \nu p_{1/2})$ character, as expected.  For the neutron-rich isotopes $^{72,74}$Cu, the ground-state spin was measured to be $I=2$ \cite{Fla10}, and from the measured negative sign of the magnetic moment it was concluded that the ground state must have a dominant $(\pi f_{5/2}\otimes \nu g_{9/2})$ configuration.  Indeed, the empirical magnetic moment of this configuration is in agreement with the observed value (see Fig.~\ref{empiricalgfactors}).

In $^{68,70}$Cu the experimental $g$-factors of the $3^-$ and $6^-$ states are consistent with a proton in the $2p_{3/2}$ level coupled to a neutron in the $g_{9/2}$ orbit (compare $\mu_{\rm emp}(3^-)= -2.85$ and $ \mu_{\rm emp}(6^-)= +1.44$ to the experimental values in Table \ref{tabmuQ2}). This configuration gives rise to a (3,4,5,6)$^-$ multiplet, where the $6^-$ state forms the ground state in $^{70}$Cu and is isomeric in $^{64-68}$Cu, due to excitation of a neutron across the $N=40$ shell gap.  The energies of these isomeric states are thus a good probe for the strength of the $N=40$ shell gap.  In Fig.~\ref{evenCulevels} we compare the lowest experimental energy levels with the calculated level schemes for both interactions.  Due to the very high level density, especially in the neutron-rich isotopes, only the lowest experimental levels and the relevant calculated ones are shown for clarity.

With the JUN45 interaction, the energy of the $6^-$ level in $^{68}$Cu is calculated about 300 keV below the observed value.  In $^{66}$Cu this level is calculated even 500 keV below the experimental value (at 661 keV, compared to 1154 keV experimentally (not shown in figure)). This supports the assumption that the $N=40$ gap is too small in the JUN45 interaction.  The energies of the $6^-$ levels calculated with jj44b are respectively 826 and 899 keV, which are 100 keV above and 250 keV below the experimental level energies in $^{68,66}$Cu.

Note that for most of the odd-odd isotopes neither theory predicts the correct level to be the ground state. In order to compare the calculated moments to the experimental ones, we have taken the calculated values from the lowest level with the correct spin (indicated by a bold line in Fig.~\ref{evenCulevels}). For $^{72}$Cu the experimental moments were best reproduced with the JUN45 interaction by the moments for the second $2^-$ level \cite{Fla10}. This illustrates the strong sensitivity of nuclear moments to the exact composition of the wave function.  That level has a dominant ($\pi f_{5/2} \otimes \nu g_{9/2}$) configuration, while the lowest $2^-$ level has a dominant $(\pi p_{3/2} \otimes \nu g_{9/2}, \sigma=3)$ configuration, with moments that do not agree with the observed value \cite{Fla10}.  The fact that the $2_2^-$ level at 645 keV is the real ground state, suggests that in the JUN45 interaction the effective single particle energy of the $\pi f_{5/2}$ level is probably too high in $^{71}$Cu.

\begin{figure}[h]
\includegraphics[scale=0.35]{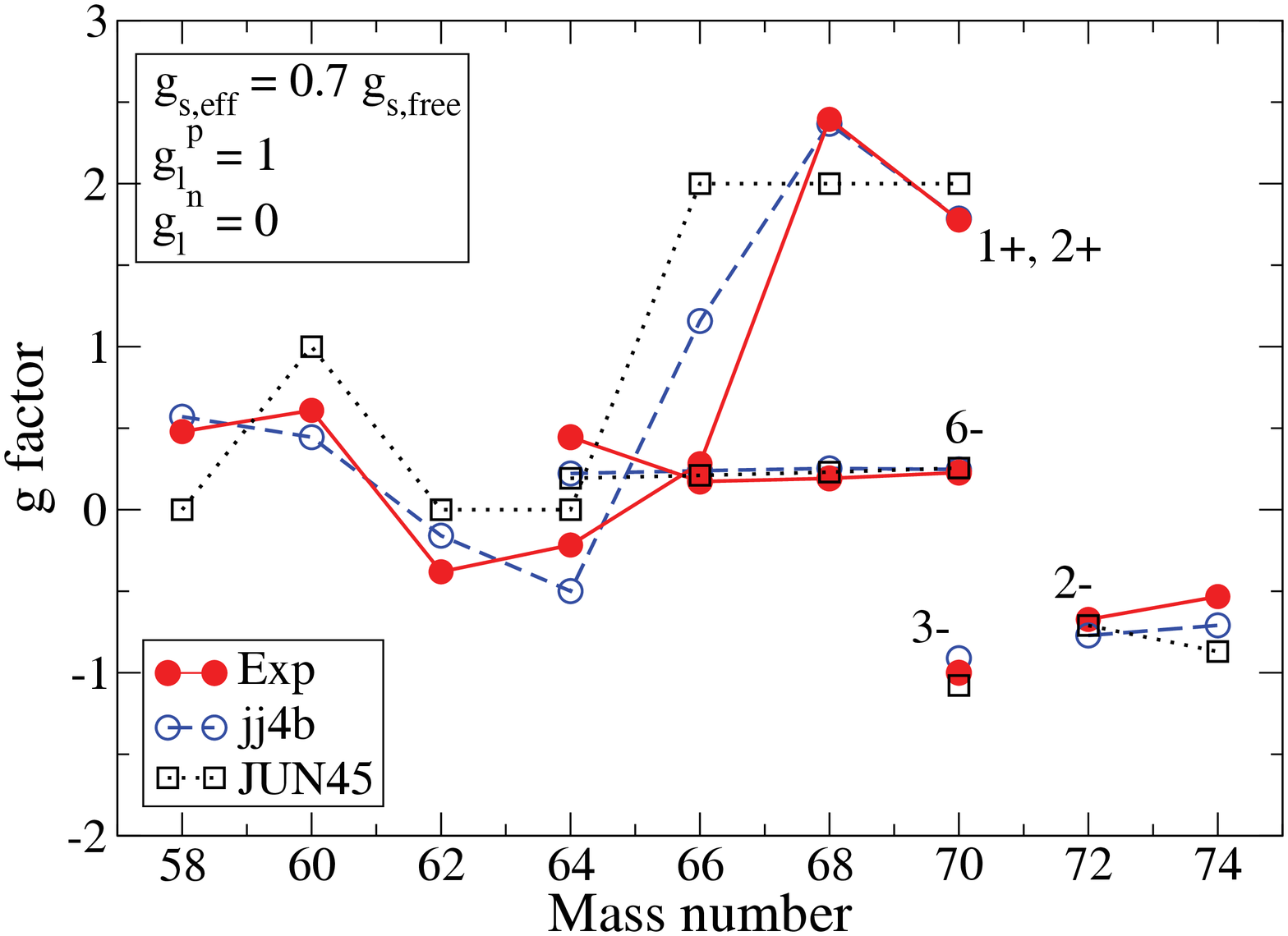}
\caption{(color online) Experimental $g$-factors for the odd-odd Cu isotopes (full dots) compared to calculations. The nuclear spin is added to distinguish between the different isomers in $^{64-70}$Cu (experimental values from Table \ref{tabmuQ2}). Good agreement is obtained, except for the $^{66}$Cu 1$^+$ ground state.}
\label{evengfactors}
\end{figure}

In Fig.~\ref{evengfactors} the measured $g$-factors are compared to the results from calculations with the jj44b and JUN45 interactions.  The jj44b interaction is very successful in predicting the $g$-factor trend for the odd-odd Cu isotopes, both for gs and isomeric states.  The agreement with the JUN45 results is also reasonable.  Only for the $^{66}$Cu ground state, the two theories significantly overestimate the observed value.  As can be concluded from the empirical $g$-factors in Fig.~\ref{empiricalgfactors}, this is because a too large $\nu p_{1/2}$ contribution is predicted to occur in the wave function. Such a contribution is certainly present, but not as much as given by the shell-model calculations.

\begin{figure}[h]
\includegraphics[scale=0.35]{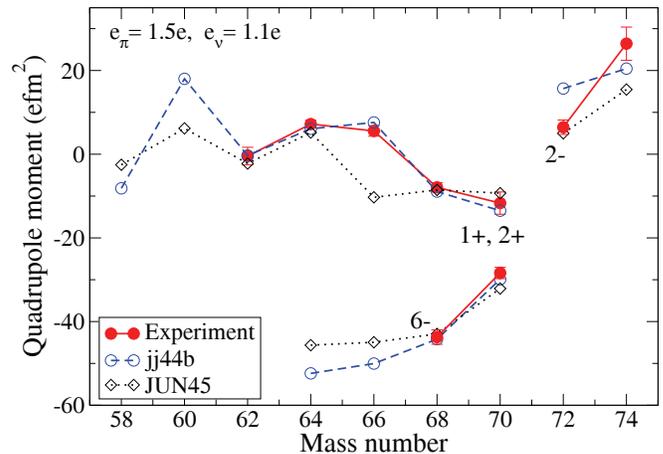}
\caption{(color online) The experimental quadrupole moments for the odd-odd Cu isotopes compared with calculations \cite{Rag89}.}
\label{evenquadrupole}
\end{figure}

Finally, we compare the quadrupole moments of the odd-odd Cu isotopes with the calculated values from both models, as shown in Fig.~ \ref{evenquadrupole}. The same effective charges have been used as for the odd-A Cu isotopes.  The jj44b interaction successfully reproduces all quadrupole moments, except for those of the $2^-$ levels in $^{72,74}$Cu where the deviation is somewhat larger.  The JUN45 interaction calculates the quadrupole moments fairly well, but it gives a wrong sign and too large a magnitude for the quadrupole moment of the $^{66}$Cu gs.  This is in line with the $g$-factor trend, where it was concluded that the $\nu p_{1/2}$ occupation in $^{66}$Cu is overestimated in this calculation.  An increased $\nu p_{1/2}$ occupation allows for enhanced neutron pairing correlations in the $\nu p_{3/2}f_{5/2}$ levels and so for extra collectivity, leading to too large a quadrupole moment.

\section{\label{summary}Summary}

In summary, the technique of collinear laser spectroscopy in combination with the ISCOOL buncher, has been successfully applied to determine the spin, magnetic and quadrupole moments of the ground states and long-lived isomeric states in the $^{61-75}$Cu isotopes with some yields as low as 10$^4$ pps. The $g$-factors and quadrupole moments of the odd-A Cu isotopes show an apparent magic behavior at $N=40$, which is strongly related to the parity change between the $pf$ shell orbits and the $g_{9/2}$ level.  Therefore this magic behavior cannot be interpreted only in terms of the energy gap at $N=40$.

The experimental results have been compared to large-scale shell-model calculations starting from a $^{56}$Ni core, using two effective shell-model interactions with protons and neutrons in the $f_{5/2}pg_{9/2}$ model space.  On both sides of $N=40$, the calculations overestimate the measured magnetic moments of the odd-A Cu isotopes. Because spin-flip excitations of the type $(f_{7/2}^{-1}f_{5/2})_{1^+}$ have a very strong influence on the magnetic moment of a state, even if this configuration contributes only one percent to the wave function \cite{Geo02}, the overestimated $g$-factors are an indication that excitations across $Z$, $N=28$ should be included in the model space. The collectivity and the onset of core polarization between $N=28$ and $N=50$ has been probed by the quadrupole moments of the odd-A Cu ground states.  No sign of an increased collectivity in $^{71,73,75}$Cu as compared to $^{63,65,67}$Cu has been observed, contrary to what was concluded from B(E2) measurements in the Ni isotopes. Extending quadrupole moment measurements towards $^{57}$Cu and $^{79}$Cu could provide more information about the softness/stiffness of the $^{56}$Ni and $^{78}$Ni cores. Comparison of the experimental core quadrupole moments with the calculated ones shows that the jj44b interaction correctly reproduces the observed core polarization when moving away from $N=40$, while the JUN45 interaction systematically underestimates the onset of collectivity.

The moments of the odd-odd Cu isotopes are very sensitive to the proton-neutron interaction and configuration mixing, and provide a more severe test to the calculations.  The jj44b interaction reproduces very well (within a few \%) all of the observed magnetic and quadrupole moments of odd-odd Cu isotopes, while for the JUN45 interaction less, but reasonable agreement was found.  Only for the $^{66}$Cu gs both models fail to reproduce correctly the data.  Note, however, that the level with moments agreeing best with experimental values, is the lowest calculated level with the correct spin. This is not always the calculated ground state.  In $^{72}$Cu it is even the second $2^-$ level that agrees with the observed gs moments calculated with JUN45. Thus, through the measured moments, a particular configuration can be assigned to the gs wave functions.

\acknowledgements
This work has been supported by the German Ministry for Education and Research (BMBF) under Contract No. 06MZ9178I, the UK Science and Technology Facilities Council (STFC), the FWO-Vlaanderen (Belgium), EU Sixth Framework through No. Eurons-506065, BriX IAP Research Program No. P6/23 (Belgium), the Max-Planck Society, NSF grant PHY-0758099 and the Helmholtz Association of German Research Centres(VH-NG-037 and VH-NG-148). M. Kowalska was supported by the EU (MEIF-CT-2006-042114). We would like to thank the ISOLDE technical group for their support and assistance during this project.


\newpage

\begin{thebibliography}{45}
\expandafter\ifx\csname natexlab\endcsname\relax\def\natexlab#1{#1}\fi
\expandafter\ifx\csname bibnamefont\endcsname\relax
  \def\bibnamefont#1{#1}\fi
\expandafter\ifx\csname bibfnamefont\endcsname\relax
  \def\bibfnamefont#1{#1}\fi
\expandafter\ifx\csname citenamefont\endcsname\relax
  \def\citenamefont#1{#1}\fi
\expandafter\ifx\csname url\endcsname\relax
  \def\url#1{\texttt{#1}}\fi
\expandafter\ifx\csname urlprefix\endcsname\relax\def\urlprefix{URL }\fi
\providecommand{\bibinfo}[2]{#2}
\providecommand{\eprint}[2][]{\url{#2}}

\bibitem[{\citenamefont{{W.F. Mueller \textit{et al.}}}(1999)}]{Mue99}
\bibinfo{author}{\bibnamefont{{W.F. Mueller \textit{et al.}}}},
  \bibinfo{journal}{Phys. Rev. Lett} \textbf{\bibinfo{volume}{83}},
  \bibinfo{pages}{3613} (\bibinfo{year}{1999}).

\bibitem[{\citenamefont{{O. Sorlin \textit{et al.}}}(2002)}]{Sor02}
\bibinfo{author}{\bibnamefont{{O. Sorlin \textit{et al.}}}},
  \bibinfo{journal}{Phys. Rev. Lett.} \textbf{\bibinfo{volume}{88}},
  \bibinfo{pages}{092501} (\bibinfo{year}{2002}).

\bibitem[{\citenamefont{{K. Langanke \textit{et al.}}}(2003)}]{Lan03}
\bibinfo{author}{\bibnamefont{{K. Langanke \textit{et al.}}}},
  \bibinfo{journal}{Phys. Rev. C} \textbf{\bibinfo{volume}{67}},
  \bibinfo{pages}{044314} (\bibinfo{year}{2003}).

\bibitem[{\citenamefont{{C. Gu\`{e}naut \textit{et al.}}}(2007)}]{Gue07}
\bibinfo{author}{\bibnamefont{{C. Gu\`{e}naut \textit{et al.}}}},
  \bibinfo{journal}{Phys. Rev. C} \textbf{\bibinfo{volume}{75}},
  \bibinfo{pages}{044303} (\bibinfo{year}{2007}).

\bibitem[{\citenamefont{{O. Kenn \textit{et al.}}}(2001)}]{Ken01}
\bibinfo{author}{\bibnamefont{{O. Kenn \textit{et al.}}}},
  \bibinfo{journal}{Phys. Rev. C} \textbf{\bibinfo{volume}{63}},
  \bibinfo{pages}{064306} (\bibinfo{year}{2001}).

\bibitem[{\citenamefont{{T.E. Cocolios \textit{et al.}}}(2009)}]{Coc09}
\bibinfo{author}{\bibnamefont{{T.E. Cocolios \textit{et al.}}}},
  \bibinfo{journal}{Phys. Rev. Lett.} \textbf{\bibinfo{volume}{103}},
  \bibinfo{pages}{102501} (\bibinfo{year}{2009}).

\bibitem[{\citenamefont{{M. Honma \textit{et al.}}}(2002)}]{Hon02}
\bibinfo{author}{\bibnamefont{{M. Honma \textit{et al.}}}},
  \bibinfo{journal}{Phys. Rev. C} \textbf{\bibinfo{volume}{65}},
  \bibinfo{pages}{061301(R)} (\bibinfo{year}{2002}).

\bibitem[{\citenamefont{{M. Honma \textit{et al.}}}(2004)}]{Hon04}
\bibinfo{author}{\bibnamefont{{M. Honma \textit{et al.}}}},
  \bibinfo{journal}{Phys. Rev. C} \textbf{\bibinfo{volume}{69}},
  \bibinfo{pages}{034355} (\bibinfo{year}{2004}).

\bibitem[{\citenamefont{{P. Raghavan}}(1989)}]{Rag89}
\bibinfo{author}{\bibnamefont{{P. Raghavan}}}, \bibinfo{journal}{At. Data Nucl.
  Data Tables} \textbf{\bibinfo{volume}{42}}, \bibinfo{pages}{189}
  (\bibinfo{year}{1989}).

\bibitem[{\citenamefont{{J. Rikovska \textit{et al.}}}(2000)}]{Rik00}
\bibinfo{author}{\bibnamefont{{J. Rikovska \textit{et al.}}}},
  \bibinfo{journal}{Phys. Rev. Lett.} \textbf{\bibinfo{volume}{85}},
  \bibinfo{pages}{1392} (\bibinfo{year}{2000}).

\bibitem[{\citenamefont{{J. Rikovska and N. J. Stone}}(2000)}]{Rik00b}
\bibinfo{author}{\bibnamefont{{J. Rikovska and N. J. Stone}}},
  \bibinfo{journal}{Hyp. Interact.} \textbf{\bibinfo{volume}{129}},
  \bibinfo{pages}{131} (\bibinfo{year}{2000}).

\bibitem[{\citenamefont{{N.J. Stone}}(2008)}]{Sto08}
\bibinfo{author}{\bibnamefont{{N.J. Stone}}}, \bibinfo{journal}{Phys. Rev. C}
  \textbf{\bibinfo{volume}{77}}, \bibinfo{pages}{014315}
  (\bibinfo{year}{2008}).

\bibitem[{\citenamefont{{G. Georgiev \textit{et al.}}}(2002)}]{Geo02}
\bibinfo{author}{\bibnamefont{{G. Georgiev \textit{et al.}}}},
  \bibinfo{journal}{J. Phys. G} \textbf{\bibinfo{volume}{28}},
  \bibinfo{pages}{2993} (\bibinfo{year}{2002}).

\bibitem[{\citenamefont{{K.T. Flanagan \textit{et al.}}}(2009)}]{Fla09}
\bibinfo{author}{\bibnamefont{{K.T. Flanagan \textit{et al.}}}},
  \bibinfo{journal}{Phys. Rev. Lett.} \textbf{\bibinfo{volume}{103}},
  \bibinfo{pages}{142501} (\bibinfo{year}{2009}).

\bibitem[{\citenamefont{{T. Otsuka \textit{et al.}}}(2005)}]{Ots05}
\bibinfo{author}{\bibnamefont{{T. Otsuka \textit{et al.}}}},
  \bibinfo{journal}{Phys. Rev. Lett.} \textbf{\bibinfo{volume}{95}},
  \bibinfo{pages}{232502} (\bibinfo{year}{2005}).

\bibitem[{\citenamefont{{T. Otsuka \textit{et al.}}}(2010)}]{Ots10}
\bibinfo{author}{\bibnamefont{{T. Otsuka \textit{et al.}}}},
  \bibinfo{journal}{Phys. Rev. Lett.} \textbf{\bibinfo{volume}{104}},
  \bibinfo{pages}{012501} (\bibinfo{year}{2010}).

\bibitem[{\citenamefont{{M. Honma \textit{et al.}}}(2009)}]{Hon09}
\bibinfo{author}{\bibnamefont{{M. Honma \textit{et al.}}}},
  \bibinfo{journal}{Phys. Rev. C} \textbf{\bibinfo{volume}{80}},
  \bibinfo{pages}{064323} (\bibinfo{year}{2009}).

\bibitem[{\citenamefont{{K. Sieja and F. Nowacki}}(2010)}]{Sie10}
\bibinfo{author}{\bibnamefont{{K. Sieja and F. Nowacki}}},
  \bibinfo{journal}{Phys. Rev. C} \textbf{\bibinfo{volume}{81}},
  \bibinfo{pages}{061303} (\bibinfo{year}{2010}).

\bibitem[{\citenamefont{{B. Cheal and K.T. Flanagan}}(2010)}]{Che10b}
\bibinfo{author}{\bibnamefont{{B. Cheal and K.T. Flanagan}}},
  \bibinfo{journal}{J. Phys. G} \textbf{\bibinfo{volume}{37}},
  \bibinfo{pages}{113101} (\bibinfo{year}{2010}).

\bibitem[{\citenamefont{{H. Franberg \textit{et al.}}}(2008)}]{Fra08}
\bibinfo{author}{\bibnamefont{{H. Franberg \textit{et al.}}}},
  \bibinfo{journal}{Nucl. Inst. Meth. B} \textbf{\bibinfo{volume}{266}},
  \bibinfo{pages}{4502} (\bibinfo{year}{2008}).

\bibitem[{\citenamefont{{E. Man\'{e} \textit{et al.}}}(2009)}]{Man09}
\bibinfo{author}{\bibnamefont{{E. Man\'{e} \textit{et al.}}}},
  \bibinfo{journal}{Eur. Phys. J. A.} \textbf{\bibinfo{volume}{42}},
  \bibinfo{pages}{503} (\bibinfo{year}{2009}).

\bibitem[{\citenamefont{{U. K\"{o}ster \textit{et al.}}}(2000)}]{Kos00}
\bibinfo{author}{\bibnamefont{{U. K\"{o}ster \textit{et al.}}}},
  \bibinfo{journal}{Nucl. Instr. Meth. B} \textbf{\bibinfo{volume}{160}},
  \bibinfo{pages}{528} (\bibinfo{year}{2000}).

\bibitem[{\citenamefont{{A. Krieger \textit{et al.}}}(2010)}]{Kri10}
\bibinfo{author}{\bibnamefont{{A. Krieger \textit{et al.}}}},
  \bibinfo{journal}{NIM A}  (\bibinfo{year}{2010}), \bibinfo{note}{submitted}.

\bibitem[{\citenamefont{{J. Van Roosbroeck \textit{et al.}}}(2004)}]{Van04}
\bibinfo{author}{\bibnamefont{{J. Van Roosbroeck \textit{et al.}}}},
  \bibinfo{journal}{Phys. Rev. Lett.} \textbf{\bibinfo{volume}{92}},
  \bibinfo{pages}{112501} (\bibinfo{year}{2004}).

\bibitem[{\citenamefont{{I. Stefanescu \textit{et al.}}}(2007)}]{Ste07}
\bibinfo{author}{\bibnamefont{{I. Stefanescu \textit{et al.}}}},
  \bibinfo{journal}{Phys. Rev. Lett.} \textbf{\bibinfo{volume}{98}},
  \bibinfo{pages}{122701} (\bibinfo{year}{2007}).

\bibitem[{\citenamefont{{Y. Ting and H. Lew}}(1957)}]{Tin57}
\bibinfo{author}{\bibnamefont{{Y. Ting and H. Lew}}}, \bibinfo{journal}{Phys.
  Rev.} \textbf{\bibinfo{volume}{105}}, \bibinfo{pages}{581}
  (\bibinfo{year}{1957}).

\bibitem[{\citenamefont{{J. Ney}}(1966)}]{Ney66}
\bibinfo{author}{\bibnamefont{{J. Ney}}}, \bibinfo{journal}{Z. Phys.}
  \textbf{\bibinfo{volume}{196}}, \bibinfo{pages}{53} (\bibinfo{year}{1966}).

\bibitem[{\citenamefont{{K.T. Flanagan \textit{et al.}}}(2010)}]{Fla10}
\bibinfo{author}{\bibnamefont{{K.T. Flanagan \textit{et al.}}}},
  \bibinfo{journal}{Phys. Rev. C.} \textbf{\bibinfo{volume}{82}}, \bibinfo{pages}{041302}  (\bibinfo{year}{2010}).

\bibitem[{\citenamefont{{P.R. Locher}}(1974)}]{Loc74}
\bibinfo{author}{\bibnamefont{{P.R. Locher}}}, \bibinfo{journal}{Phys. Rev. B}
  \textbf{\bibinfo{volume}{10}}, \bibinfo{pages}{801} (\bibinfo{year}{1974}).

\bibitem[{\citenamefont{Data}()}]{NNDC}
\bibinfo{author}{\bibfnamefont{E.~N.~S.} \bibnamefont{Data}},
  \bibinfo{note}{http://www.nndc.bnl.gov/}.

\bibitem[{\citenamefont{{L. Weissman \textit{et al.}}}(2002)}]{Wei02}
\bibinfo{author}{\bibnamefont{{L. Weissman \textit{et al.}}}},
  \bibinfo{journal}{Phys. Rev. C} \textbf{\bibinfo{volume}{65}},
  \bibinfo{pages}{024315} (\bibinfo{year}{2002}).

\bibitem[{\citenamefont{{K. Blaum \textit{et al.}}}(2004)}]{Blau04}
\bibinfo{author}{\bibnamefont{{K. Blaum \textit{et al.}}}},
  \bibinfo{journal}{Europhys. Lett.} \textbf{\bibinfo{volume}{67}},
  \bibinfo{pages}{586} (\bibinfo{year}{2004}).

\bibitem[{\citenamefont{{S. Gheysen \textit{et al.}}}(2004)}]{Ghe04}
\bibinfo{author}{\bibnamefont{{S. Gheysen \textit{et al.}}}},
  \bibinfo{journal}{Phys. Rev. C} \textbf{\bibinfo{volume}{69}},
  \bibinfo{pages}{064310} (\bibinfo{year}{2004}).

\bibitem[{\citenamefont{{V.V. Golovko \textit{et al.}}}(2004)}]{Gol04}
\bibinfo{author}{\bibnamefont{{V.V. Golovko \textit{et al.}}}},
  \bibinfo{journal}{Phys. Rev. C} \textbf{\bibinfo{volume}{70}},
  \bibinfo{pages}{014312} (\bibinfo{year}{2004}).

\bibitem[{\citenamefont{{T. E. Cocolios \textit{et al.}}}(2010)}]{Coc10}
\bibinfo{author}{\bibnamefont{{T. E. Cocolios \textit{et al.}}}},
  \bibinfo{journal}{Phys. Rev. C} \textbf{\bibinfo{volume}{81}},
  \bibinfo{pages}{014314} (\bibinfo{year}{2010}).

\bibitem[{\citenamefont{{D. Verney \textit{et al.}}}(2007)}]{Ver07}
\bibinfo{author}{\bibnamefont{{D. Verney \textit{et al.}}}},
  \bibinfo{journal}{Phys. Rev. C} \textbf{\bibinfo{volume}{76}},
  \bibinfo{pages}{054312} (\bibinfo{year}{2007}).

\bibitem[{\citenamefont{{G. Neyens}}(2003)}]{Ney03}
\bibinfo{author}{\bibnamefont{{G. Neyens}}}, \bibinfo{journal}{Rep. Prog.
  Phys.} \textbf{\bibinfo{volume}{66}}, \bibinfo{pages}{633}
  (\bibinfo{year}{2003}).

\bibitem[{\citenamefont{{S. Franchoo \textit{et al.}}}(2001)}]{Fra01}
\bibinfo{author}{\bibnamefont{{S. Franchoo \textit{et al.}}}},
  \bibinfo{journal}{Phys. Rev. C} \textbf{\bibinfo{volume}{64}},
  \bibinfo{pages}{054308} (\bibinfo{year}{2001}).

\bibitem[{\citenamefont{{I. Stefanescu \textit{et al.}}}(2008)}]{Ste08}
\bibinfo{author}{\bibnamefont{{I. Stefanescu \textit{et al.}}}},
  \bibinfo{journal}{Phys.\ Rev. Lett.} \textbf{\bibinfo{volume}{100}},
  \bibinfo{pages}{112502} (\bibinfo{year}{2008}).

\bibitem[{\citenamefont{{I. Stefanescu \textit{et al.}}}(2009)}]{Ste09}
\bibinfo{author}{\bibnamefont{{I. Stefanescu \textit{et al.}}}},
  \bibinfo{journal}{Phys. Rev. C} \textbf{\bibinfo{volume}{79}},
  \bibinfo{pages}{044325} (\bibinfo{year}{2009}).

\bibitem[{\citenamefont{{J. M. Daugas \textit{et al.}}}(2010)}]{Dau10}
\bibinfo{author}{\bibnamefont{{J. M. Daugas \textit{et al.}}}},
  \bibinfo{journal}{Phys. Rev. C} \textbf{\bibinfo{volume}{81}},
  \bibinfo{pages}{034304} (\bibinfo{year}{2010}).

\bibitem[{\citenamefont{{B. Cheal \textit{et al.}}}(2010)}]{Che10}
\bibinfo{author}{\bibnamefont{{B. Cheal \textit{et al.}}}},
  \bibinfo{journal}{Phys. Rev. Lett.} \textbf{\bibinfo{volume}{104}},
  \bibinfo{pages}{252502} (\bibinfo{year}{2010}).

\bibitem[{\citenamefont{{O. Perru \textit{et al.}}}(2006)}]{Per06}
\bibinfo{author}{\bibnamefont{{O. Perru \textit{et al.}}}},
  \bibinfo{journal}{Phys. Rev. Lett.} \textbf{\bibinfo{volume}{96}},
  \bibinfo{pages}{232501} (\bibinfo{year}{2006}).

\bibitem[{\citenamefont{{N. Aoi \textit{et al.}}}(2010)}]{Aoi10}
\bibinfo{author}{\bibnamefont{{N. Aoi \textit{et al.}}}},
  \bibinfo{journal}{Phys. Lett. B} \textbf{\bibinfo{volume}{692}},
  \bibinfo{pages}{302} (\bibinfo{year}{2010}).

\bibitem[{\citenamefont{{J.-C. Thomas \textit{et al.}}}(2006)}]{JCT06}
\bibinfo{author}{\bibnamefont{{J.-C. Thomas \textit{et al.}}}},
  \bibinfo{journal}{Phys. Rev. C} \textbf{\bibinfo{volume}{74}},
  \bibinfo{pages}{054309} (\bibinfo{year}{2006}).

\end{thebibliography}
\end{document}